\documentclass[runningheads]{llncs}
\usepackage[T1]{fontenc}
\usepackage{amsmath,amssymb}
\usepackage{graphicx}
\usepackage[all]{xy}
\usepackage{url}
\usepackage{macros}
\usepackage{todonotes}
\usepackage{framed}
\usepackage{mathtools,
		     stmaryrd,
	     	     bussproofs,
       		     bbm,
            	     tikz-cd,
            	     mathrsfs}
\usepackage{cite}

\usepackage[createShortEnv,commandRef=Cref,conf={no link to proof}]{proof-at-the-end}
\pratendSetGlobal{text proof={Proof of \string\pratendRef{thm:prAtEnd\pratendcountercurrent} on \string\cpageref{thm:prAtEnd\pratendcountercurrent}}}
\usepackage[hidelinks]{hyperref}
\usepackage[nameinlink,
		    capitalize]{cleveref}
\crefname{theorem}{Theorem}{Theorems}
\Crefname{theorem}{Theorem}{Theorems}
\crefname{definition}{Definition}{Definitions}
\Crefname{definition}{Definition}{Definitions}
\crefname{example}{Example}{Examples}
\Crefname{example}{Example}{Examples}
\crefname{remark}{Remark}{Remarks}
\Crefname{remark}{Remark}{Remarks}
\crefname{equation}{Equation}{Equations}
\Crefname{equation}{Equation}{Equations}

\spnewtheorem*{proofsketch}{Proof sketch of Theorem 4}{\itshape}{\normalfont}
\begin{document}
\title{Final coalgebras over $R$-modules}
\titlerunning{Final coalgebras over $R$-modules}

\author{
Ois\'{i}n Flynn-Connolly
 \and 
Henning Basold 
 \and
Chase Ford}
\authorrunning{O. Flynn-Connolly, H. Basold, C. Ford}
\institute{LIACS,
Leiden University, Leiden, The Netherlands\\
\email{\{o.c.flynn-connolly, h.basold, m.c.ford\}@liacs.leidenuniv.nl}}
\maketitle             
\begin{abstract}
  Final coalgebras and the coinduction principle for coalgebras over $R$-modules, where $R$ is a commutative ring, are essential in determining the denotational behaviour of various types of linear weighted automata.
  The goal of this paper is to provide lower and upper bounds on the ordinals that are needed to construct final coalgebras over $R$-modules as the limit of a transfinite sequence of functor iterations.
  We give an example of a binary intersection preserving finitary functor on the category of abelian groups, which is the category of modules over integers, that fails to converge in $\omega \cdot 2$ steps, thus settling an open problem of Ad\'{a}mek, Milius, and Moss.
  Then we relate this to the failure of the Artinian property of the integers and show that all finitary functors on $R$-modules for a Noetherian ring $R$ with Krull dimension $n$, which preserve countable directed intersections and a generalised notion of monomorphism, converge in $\omega \cdot (n + 2)$ steps.
  We also establish how the connection between Aczel-Mendler bisimulation and behavioural equivalence is mediated by the functor $\operatorname{Ext}^1$.
  These results show that homological algebra tools can play an important role in coalgebra.
\end{abstract}
\section{Introduction}
\label{sec:intro}

Modules over a commutative ring $R$ play an important role in reasoning about the behaviour of linear sequential machines~\cite{eilenberg74}, weighted automata and their coalgebraic models~\cite{Basold-LinearCiruits,bonchi12,Boreale09,Milius10:SoundCompleteCalculus}.
Final coalgebras and the coinduction principle for coalgebras over $R$-modules are then essential in determining the denotational behaviour of various types of linear weighted automata.
For instance, the existence of a final coalgebra leads to very natural equational definition and proof principles~\cite{Rutten00,HKR14:SDE,Rutten05:CoindCalcStreams,ruttenTutorialCoinductiveStream2005}.
A central question is then under what conditions final coalgebras exist.
There are various well-known constructions, in particular final coalgebras can be constructed as limit of a well-founded sequence of functor iterations, which provides control over the shape of the final coalgebra.
Famously, Worrell~\cite{Worrell99,Worrell05} has shown that many interesting functors need, however, transfinite iterations in order for this sequence to converge to a final coalgebra.
In recent work, the convergence result of Worrell was extended to a broad class of categories and functors, where the iteration bound can be determined~\cite{AdamekEA25}.
However, it was left open whether the conditions apply to abelian groups, which are the same as modules over the integers $\Z$.

There has been no previous detailed study of final coalgebras or coinduction in abelian groups because the functors of primary interest to algebraists are primarily additive, which in particular preserve zero objects.
Recall that a zero object is simultaneously an initial and a terminal object in a category.
In the category $\RMod$ of $R$-modules and linear maps, the modules with $0$ as its only element is a zero object.
For commutative groups, this instantiates to the singleton group.
Since the construction of a final coalgebra for a functor $F$ proceeds by iterating $F$ on the terminal object, the construction trivially converges if $F$ is additive.
Thus the final coalgebra is a zero object and behavioural equivalence trivialises.
In this article we investigate further the properties of \emph{non-additive functors}, which include models for linear weighted automata, in abelian groups and $R$-modules.
We set out in this paper to provide convergence bounds for the final coalgebra construction for functors on $\RMod$.

The usual procedure for finding final coalgebras for a functor $F \from \CC \to \CC$ on a category $\CC$ is by constructing the $\op{\omega}$-chain that starts with the terminal object $1$ and iterates the functor $F$, where $m \from F1 \to 1$ is the terminal map:
 \begin{equation}
 \label{omegach}
\begin{tikzcd}
\cdots \rar{F^3m} & F^31 \rar{F^2m} & F^21 \rar{Fm} & F1  \rar{m} & 1
\end{tikzcd}
\end{equation}
If the functor $F$ preserves $\op{\omega}$-limits, then this chain \emph{converges} to a final coalgebra~\cite{Adamek03}.
However, if the functor does not preserve those limits, then the process requires transfinitely many steps and may not work at all if the functor is unbounded.
For example, applications of the power set functor always strictly increase the cardinality and cannot have a fixed point.
In a remarkable result, Worrell~\cite{Worrell99,Worrell05} showed that all finitary, that is to say, filtered colimit preserving, functors on the category $\Set$ of sets and maps are guaranteed to converge in at most $\omega \cdot 2$ steps.
He also showed that for some functors, like the finitary power set functor, this bound on steps is actually obtained.

In recent work, Ad\'{a}mek, Milius, and Moss~\cite{AdamekEA25} generalised Worrell's convergence results from the category of sets to locally finitely presentable (lfp) categories, where the finitely presentable objects satisfy a descending chain condition (DCC), and functors preserving binary intersections.
For modules over a ring, the DCC property is connected to the Artinian property of the ring, so the category of abelian groups, which is isomorphic to the category of $\Z$-modules, is the simplest example of a non-DCC lfp category.
In this paper, we explicitly construct a functor $U$ that does not converge in $\omega \cdot 2$ steps.
Our construction exploits the fact that the subcategory of torsion groups, while closed under colimits, is not closed under limits in the category of abelian groups.
Non-torsion groups can be detected using tensor products with $\Q$, and we devise our functor to be a direct sum of functors $F_p\oplus T,$ where the $T$-component only becomes non-zero beyond $\omega$ steps, when non-torsion groups start to appear.
This prevents the construction from converging in $\omega \cdot 2$ steps and thus resolves the problem left open by Ad\'{a}mek, Milius, and Moss~\cite{AdamekEA25}.

Having thus exhibited the rather unfortunate pathologies of $\RMod$ from a coalgebraic perspective, we turn to finding tractable criteria for determining convergence of final coalgebra sequence in $\omega \cdot n$ many steps.
For this, we work with Noetherian rings $R$ of (finite) \emph{Krull dimension} $n$, which is the length of the longest chain of proper prime ideals contained in $R$.
We prove that in this case for a finitary functor on $\RMod$ that preserves, both, countable directed intersections of monomorphisms and a refinement of monomorphism, the chain construction converges in $\omega \cdot (n+2)$ steps to its final coalgebra.
In the case of abelian groups, this guarantees convergence in $\omega\cdot 3$ steps to suitably well behaved functors.

Along the way, we also exhibit novel behaviour here not seen in $\Set$ by showing that the equivalence between Aczel-Mendler bisimulation and behavioural equivalence is mediated by the $\operatorname{Ext}^1$-functor, not just weak pullback preservation\cite{Rutten00,Staton2011-Bisim}.
We believe that it shows the potential for laying deeper connections between coalgebra and homological algebra.

\paragraph{Related work.}
The $p$-adic numbers, which were the main inspiration for our counterexample, have been studied from a coalgebraic perspective in  \cite{Bhattacharya15}, and the fractal-like nature of the functor $F_p$ is somewhat reminiscent of the self-similarity work appearing in \cite{leinster11, schmid25}. However such connections appear to be merely surface level as the authors there are more interested in metric spaces.

Of much greater relevance to our work is the literature on \emph{linear weighted automata} \cite{Boreale09, bonchi12}. This was originally worked out in vector spaces and subsequently generalised to semimodules over semirings \cite{Bhaduri23}. Learning algorithms for linear weighted automata over commutative rings have been developed in \cite{rot20, aristote26}. Our paper sits in between these two as it has relevance to linear weighted automata valued in commutative rings, though we work with much more general functors than in either paper, and are not interested in learning algorithms. 

Aczel-Mendler bisimulation in the category of $R$-modules, among other categories, was revisited in \cite{dubut25} where it is shown that linear weighted automata functors there preserve weak pullbacks. We also briefly study bisimulation, and seem to be the first to make the connection to homological algebra observed in Proposition \ref{prop:ext}.

 Descending chain conditions were first introduced in the coalgebra literature in \cite[Section 4]{urbat20} and have been used to study learning algorithms for weighted automata in the context of rings and semirings  \cite{rot20, aristote26} and bisimulation directly in \cite{Bhaduri23}. We address a point of potential confusion here, the notion referred to \cite[Section 4]{urbat20} as \emph{Noetherian} is, in the case of modules more closely related to the classical notion of \emph{Artinian} while Noetherianity is a strictly weaker notion in this context. We have therefore adopted the convention from commutative algebra. We note that this apparently minor terminological confusion has already caused some issues in the literature, for example, the argument given for the main theorem in one paper  \cite[Theorem 8]{aristote26} appears to rely on a result that requires the Artinian property, in a setting where it fails. The insight that Artinian can be weakened to Krull dimension appears to be new to this paper.
 
Finally, the $\T_d$ discussed in Section \ref{sec:omega3} form \emph{(hereditary) torsion theories}~\cite{dickson66} which, to the best of our knowledge, have not been explicitly exploited in the study of coalgebras before. 
This relates to concept of \emph{Gabriel-Krull dimension} of an abelian category due to Gabriel~\cite{gabriel62}.

\section{Preliminaries}
\label{sec:prelim}
\pratendSetLocal{category=prelim}
We briefly gather the background material and notation on commutative algebra \cite{eisenbud95}, coalgebra~\cite{Rutten00} and $R$-modules that will be needed throughout the paper.  
We shall assume basic knowledge of category theory, see \cite{MacLane98} for a refresher on that. 
 The reader well-versed in these topics may choose to skip this section on first reading.
 The results of this section that are original are Proposition \ref{prop:ext}, Proposition \ref{prop:monopres} and along with most of the examples.

\paragraph*{Notions of finiteness for rings.}
  Let $R$ be a (commutative) ring, i.e., a set $R$ equipped with two \emph{commutative} binary operations: an addition operation $+$ and a multiplication operation $\cdot$, with identities $0$ and $1$ respectively. We require additive inverses but not necessarily multiplicative ones. Lastly, multiplication is required to distribute over addition. In other words, $R$ is a commutative monoid in abelian groups. To avoid pathologies, we always assume $R\neq 0$.

  An \emph{ideal} $I$ of a ring $R$ is a subgroup of $(R, +,0)$ such that  $IR \subseteq I$, where $IR =\setDef{i\cdot r }{i\in I, r\in R}$.
  An example is the ideal $n\Z \coloneqq \setDef{nm}{m\in \Z}$: the ideal $2\Z$ is the even numbers. Any set of elements $\setDef{r_i\in R}{ i \in \mathcal I}$ \emph{generates} an ideal $(r_i)_{i \in \mathcal I} = \setDef{\sum_{i\in \mathcal I} r_ig_i }{g_i \in R }$ of  $R$ e.g., $(2)$ is an alternative notation for $2\Z$. An ideal is \emph{prime} if it is not equal to $R$  and $p\cdot q \in I$ only if at least one of $p,q$ is an element of $I$; the notation comes from the fact that $n\Z$ is a prime ideal if and only if $n$ is zero or a prime number. A ring is a \emph{domain} if the ideal $\{0\}$ is prime i.e., it has no nontrivial zero divisors. Given an ideal, there is a notion of a \emph{quotient} $q\from R \to R/I$ determined by the usual universal property of the quotient against all maps that are zero on $I.$ The ideals (resp.\ prime ideals) of $R/I$ are in a one-to-one correspondence induced by $q$ with the ideals (resp.\ prime ideals)  of $R$ containing $I$.

In general, rings can behave quite pathologically so it is natural to consider some finiteness assumptions on them.  A ring is called \emph{Noetherian} if it satisfies the \emph{ascending chain condition}: every strictly ascending chain of ideals $I_1 \subsetneq I_2 \subsetneq I_3 \subsetneq  \cdots $ eventually terminates. The ring $\Z$, along with all fields and finite rings, is Noetherian; this is a general running assumption throughout much of commutative algebra and on the rings we consider in this work. One of the main reasons it is useful is that every ideal of a Noetherian ring has finitely many generators. A ring is called \emph{Artinian} if it satisfies the \emph{descending chain condition}: every strictly descending chain of ideals $I_1 \supsetneq I_2 \supsetneq I_3 \supsetneq  \cdots $ eventually terminates.  All fields and finite rings are Artinian. Every Artinian ring is Noetherian \cite[Theorem 2.14]{eisenbud95}, but the converse is false: starting with the integers $\Z$, as the descending chain $2\Z\supsetneq 2^2\Z \supsetneq  2^3 \Z \supsetneq  \cdots $ does not terminate. 
 
The notion of Artinian ring can be generalised further to the notion of \emph{Krull dimension}. The Krull dimension of ring $R$, denoted $\dim R$ is $n$ if the longest descending chain of prime ideals $\p_0 \supsetneq \p_1 \supsetneq \cdots  \supsetneq \p_n$ is of length $n$. For example, $\dim \Z =1 $ as the longest chain is $(0) \subsetneq p\Z$ because prime ideals do not strictly contain any other nonzero prime ideals. On the other hand, the ring of integral polynomials in one variable $\mathbb Z[x]$ has Krull dimension 2, as we have an inclusion of prime ideals $(0) \subsetneq (x)  \subsetneq (2, x)$, where $(x), (2,x)$  are the smallest ideals containing $\{x\}$ and $\{2,x\}$ respectively. This is an instance of a more general fact: if  $R$ is Noetherian and $\dim R = n$ then $\dim R[x] = n+1.$ A ring is Artinian if and only if it is Noetherian and of Krull dimension 0.

\paragraph*{$R$-modules.} 
Next we establish some basic facts about the categories of modules that will be helpful in the sequel.
\begin{definition}
  Given a ring $R$, an \emph{$R$-module} is an abelian group $(M, +)$ equipped with an $R$-action $R\times M \to M$, written as $rm$ for $(r, m) \in R \times M$, such that $r(m+n) = rm+rn$, $r(sm) = (r\cdot s)m,$ $(r+s) m = rm +sm$ and $1m = m$ for $r,s \in R$ and $m,n\in M$.
  A morphism of $R$-modules $f \colon M\to N$ is a morphism of abelian groups that is equivariant with respect to the action of $R$. We denote the category of $R$-modules formed thus by $\RMod.$
\end{definition}
\begin{example}
\label{abeliangroups}
Given an abelian group $(G, +, e)$, an element $g \in G$ and a positive integer $n \in \Z$ we define
\begin{equation*}
ng = \underbrace{g + g + \cdots + g}_{\text{n times}}, \quad 0 g = e,   \quad -ng = \underbrace{(-g) + (-g) + \cdots + (-g)}_{\text{n times}}
\end{equation*} 
This equips the set $G$ with a canonical $\Z$-module structure. Conversely, it can easily be shown that every $\Z$-module arises this way. 
This extends to an isomorphism of categories between $\Z$-modules and abelian groups.
\end{example}
The category $\RMod$ is monadic over $\Set.$
From this, it immediately follows that the category $\RMod$ has all limits and these are preserved by the forgetful functor. It is cocomplete \cite[Section 2.6]{weibel94}.
This category also has a zero object given by \emph{zero module} $0$. Explicitly, this is the group with one element $\{e\}$ equipped with the trivial action $re = e$ for all $r\in R$.

As $\RMod$ has all limits and colimits it has, in particular, products and coproducts. Finite coproducts and products coincide and are both denoted by the direct sum notation $M\oplus N$. In the infinite case, these no longer will be the same, and in that case, to maintain consistency with usual mathematical notation, we shall denote infinite products as $\prod_{i\in \mathcal I} M_i$ and infinite coproducts by $\bigoplus_{i \in \mathcal I} M_i$. Given $m \in\bigoplus_{i \in \mathcal I} M_i$, we denote by the \emph{support} of $m$
\begin{equation*}
\operatorname{supp} m \coloneqq \setDef{i \in \mathcal I}{\pi_i(m)\neq 0},
\end{equation*} note that this set is always finite. A functor $F\from \RMod \to \RMod$ that preserves finite products is said to be \emph{additive}.

The monomorphisms and epimorphisms on $\RMod$ are the injective and surjective maps on the underlying sets. The \emph{kernel} of a map $f \from M\to N$ is the submodule of $M$ carried by the set $\setDef{m\in M }{f(m) = 0}.$ The map $f$ is a monomorphism if and only if $\ker f = \{0\}.$ Dually the image of a map $f$ is a submodule of $N$ and the \emph{cokernel} $\coker{f}$ is defined to be the quotient module $N/\operatorname{Im} f.$ A map is epi if and only if its cokernel is zero.  Later in this paper, we shall use these facts to generalise the notion of epis and monos.

\paragraph*{Homological algebra.}
We recall some concepts from  homological algebra~\cite{weibel94}.
A \emph{short exact sequence (ses)} is a diagram
\begin{equation*}
0 \to N  \xrightarrow{f} E  \xrightarrow{g} M  \to 0
\end{equation*}
where the map $f\from N\to E$ is a monomorphism, the map $g \from E\to M$ is an epimorphism and the kernel of $g$ coincides with the image of $f$. An example is the \emph{split ses} $0 \to N \xrightarrow{\id\oplus 0} N\oplus M \xrightarrow{0\oplus \id} M \to 0$. An example of a non-split ses is $0 \to \Z/2\Z \to \Z/4\Z \to \Z/2\Z \to 0.$   

A module $M$ is said to be \emph{projective} if it has the weak left lifting property against surjective maps. All free modules are projective but the converse is false. The $\operatorname{Ext}^1(M, N)$ bifunctor  is of enormous importance in homological algebra and admits several descriptions. The one we shall use in this paper is the following: it is the set of short exact sequences of $R$-modules  $0 \to N \to E \to M \to 0$, modulo the equivalence generated by maps of short exact sequences that restrict to the identity on $N$ and $M.$ This set can be equipped with an $R$-module structure via the \emph{Baer sum}, which we shall not explicitly describe, except to say that the zero element is the equivalence class of the \emph{split} sequence $0 \to N \to N\oplus M \to M \to 0$. This is a bifunctor in $N$ and $M$.  It is easy to show that if $N$ is injective, then $\operatorname{Ext}^1(-, N)$ is identically 0 and similarly, if $M$ is projective then $\operatorname{Ext}^1(M, -)= 0$.

\paragraph*{Locally presentable category and finitary functors.}
 A diagram $\mathcal K \to \Cat{C}$ is \emph{directed} if $\mathcal K$ is a directed poset. A functor that  preserves directed colimits is said to be \emph{finitary}. An object $A$ of a category $\Cat{C}$ is finitely presentable if the functor $\Cat{C}(A, -) \from \Cat{C} \to  \Set$ preserves directed colimits. A category is \emph{locally finitely presentable} if it is cocomplete and has a set of finitely presentable objects such that every object is a directed colimit of objects from that set.  It is well known that the category of $\RMod$ is locally finitely presentable for every ring $R$, see \cite[3.5(2)]{ademek94}. The finitely presentable objects are then those admitting an exact sequence $R^n \to R^m \to M \to 0$ \cite{AdamekEA19}. For example,  the finitely presentable objects in $\ZMod$ are precisely those of the form $\Z^{\oplus n} \oplus \Z/p_1^{k_1}\Z \oplus \cdots \Z/p_n^{k_n}\Z $ with $p_i$ prime. When $R$ is Noetherian, it is well-known that every finitely generated module is also finitely presented.  \emph{In statements where $R$ is Noetherian, we use both of these terms interchangeably.}

 \begin{propositionE}[][all end]
\label{prop:AbFC}
Let $\mathcal D: \mathcal I \to \RMod$ be a directed diagram with colimit $M$. Then, the carriers in $\Set$ satisfy
\begin{equation*}
M \cong \coprod_{i\in \mathcal I} M_i / \sim
\end{equation*}
where $m_i \sim m_j$ if there exists $k\geq i,j$ such that $\mathcal D_{ik}(m_i) = \mathcal D_{jk}(m_j).$ The $R$-action  is given by $r\cdot [m] = [r\cdot m].$
\end{propositionE} 
\begin{proofE}
The forgetful functor from $\RMod$ to $\Set$ is precisely the map $\hom(R,-)\from \RMod \to \Set.$ Since $R$ itself is finitely presentable, this preserves all filtered colimits, so the underlying set of the filtered colimit is as claimed. It thus suffices to verify that the $R$-module structure is well-defined.

First, we show that the action of $R$ is well-defined, suppose that  $[m] = [m']$ where $m \in M_i$ and $m'\in M_j$. By construction, there exists $u \geq i, j$ such that $\mathcal D_{iu}(m) = \mathcal D_{ju}(m')$. As the transition maps are $R$-module homomorphisms, we have that $\mathcal D_{iu}(r \cdot m)  r \cdot \mathcal D_{iu}(m) = r\cdot  \mathcal D_{ju}(m') = \mathcal D_{iu}(r \cdot m'),$ and so we have proven that $[r\cdot m] = [r\cdot m'].$  We then clearly inherit the various definitional properties of the $R$-module structure on $M$ from that on $M_i$. 
 
Observe that $M$ satisfies the universal property of the colimit, as given any cocone, as set morphisms, the cocone maps factor through $M$ as it is the colimit in $\Set$. These are then $R$-module homomorphisms as the filtered colimit maps are $R$-module homomorphisms.
\end{proofE}
 
 We refer the reader to \cite{ademek94} for more details on locally presentable categories and further examples.

\paragraph*{Tensor products.}
The category~$\RMod$ carries the structure of a monoidal category, which we recall next.
Given two $R$-modules $A$ and $B$, their tensor product $A \otimes_{R} B$ is the $R$-module generated by symbols $a \otimes b$ for $a \in A, b \in B$, subject to the standard bilinearity relations:
\begin{align*}
&(a_1 + a_2) \otimes b = a_1 \otimes b + a_2 \otimes b && a \otimes (b_1 + b_2) = a \otimes b_1 + a \otimes b_2  \\
&ra \otimes  b = r(a\otimes b)= a\otimes rb
\end{align*}
for $r\in R$. This construction satisfies the universal property that homomorphisms $A \otimes B \to C$ correspond bijectively to bilinear maps $A \times B \to C$.
In the case when $R = \Z$, the third condition follows from the previous two.
\begin{example}
    The ring $R$, regarded as a module over itself, is the unit for the tensor product $\otimes_R$: for any module  $M$, we have a natural isomorphism $R \otimes_R M \cong M$. We also have that the tensor product $0 \otimes_R M = 0$.
\end{example}
\begin{example}
Tensor products give examples of finitary functors on $\RMod$ that do not preserve binary intersections. Consider $F(X) = \Z/2\Z \otimes_{\Z} X$. This does not preserve monomorphisms: given $f \from 2\Z\to \Z$, we have that $F(f)\from \Z/2\Z\to \Z/2\Z$ is the zero map, which is not a monomorphism. As it does not preserve monomorphisms, it cannot preserve binary intersections.
\end{example}
Tensor products give rise to a large class of functors which feature in homological algebra and the study of (finite-state) transition systems.
\begin{example}
 \label{ex:polynomial}
\begin{enumerate}
\item
A \emph{polynomial functor} in $\RMod$ is a functor of the form 
\begin{equation*}
F(X) \coloneqq \bigoplus_{n \geq 0} A_n \otimes X^{\otimes n}, 
\end{equation*}
where $A_n \in \RMod$.
Polynomial functors are finitary since forming tensors is a left adjoint is a left adjoint (hence preserves all colimits) and finitary functors are closed under coproducts. 
\item
We call a polynomial functor $F$ \emph{linear} if $A_i = 0$ for all $i\geq 1$, i.e. $FX = B\otimes X$ for some $B\in\RMod$; we call $F$ \emph{affine} if $FX = A \oplus( B\otimes X)$.  
Linear functors are additive while affine functors are typically not.
 \item
A \emph{linear weighted transition system} is a coalgebra for the affine functor $F(X) = R \oplus(R^{L} \otimes X)$ \cite{bonchi12, Bhaduri23}.  Here $L$ is a finite alphabet, and $R^{L}$ is the set of maps (in $\Set$) $f \from L \to R$, equipped with an $R$-module structure via that on the codomain. We have that $R^{L} \otimes X \cong X^L$ when $L$ is finite. Suppose the carrier is free, then we think of a coalgebra $\langle o, f \rangle \from X \to FX$ as assigning to every basis element $x\in X$ an output weight $o(x)\in R$ along with a transition structure $f(x)(l) = a_1 x_1 + \cdots + a_n x_n$  for every label  $l \in L$. We think of the sum as a superposition of states.
\end{enumerate}
\end{example}

We would like to highlight the interaction between tensor products and torsion in modules.
Let $R$ be a domain.
We say that $M$ is a \emph{torsion $R$-module} if for each $m\in M$, one has $r\cdot m = 0$ for some nonzero $r\in R$.
 We denote the full subcategory of torsion modules by $ \RTors.$ An example of a torsion $\Z$-module is $ \Z/ n\Z,$ the cyclic group of order $n$, as multiplication by $n$ is identically 0.
Relatedly, given an $R$-module $M$, there  is an associated ideal of $R$ called the \emph{annihilator}, and defined as $\Ann M = \setDef{r\in R}{ r\cdot m =0 \mbox{ for all } m\in M}$. For example the annihilator of the $\Z$-module $\Z/2\Z$ is $2\Z.$

 Conversely, we say that $M$ is a \emph{divisible $R$-module} if for every element $m \in M$ and every nonzero $r\in R$, there exists an element $n\in M$ such that $m = r\cdot n$.  
The rational numbers $\Q$ are an example of a divisible $\Z$-module.

\begin{propositionE}
\label{silentlemma}
    If $T$ is a torsion $R$-module and $M$ is a divisible $R$-module, then $M \otimes_{R} T = 0$.
\end{propositionE}
\begin{proofE}
    Let $t \in T$ and $d \in M$. Since $T$ is torsion, there exists a $n \in R$ such that $nt = 0$. Since $M$ is divisible, there exists a nonzero $m' \in M$ such that $d = nm'$. Then in the tensor product:
    $$
    d \otimes t = (nd') \otimes t = d' \otimes (nt) = d' \otimes 0 = 0.
    $$
    Since the pure tensors generate the module, the whole tensor product is 0.
\end{proofE}
The above lemma implies that the functor $D \otimes -$, for a divisible $R$-module $D$, kills all torsion groups. For example $\Q\otimes \Z/2\Z = 0.$

\paragraph*{Bisimulation, behavioural equivalence, and a cohomological obstruction.}
We now discuss coalgebra with a view toward $R$-modules. Given an endofunctor $F$ on a category $\C$, an \emph{$F$-coalgebra} is a pair $(A,f)$, where $A\in \C$ and $f\from A\to FA$ is a morphism. The object $FA$ is generally viewed as modelling the unfolding of some kind of process.

The collection of coalgebras forms a category, which under some assumptions, may have a terminal object $\nu F,$ which is viewed as the universe of all possible behaviours. If $F$ preserves the zero object, as any additive functor does, then its final coalgebra of $F$ is $0$. This is because the zero object is terminal in $\RMod$, and $\id \from 0 \to F(0) = 0$ is
then a coalgebra to which every coalgebra $(A, f)$ admits a unique morphism, namely the zero map.

We now turn to discussing (Aczel-Mendler) bisimulation in $\RMod$.  Below, given two morphisms $f \from A \to X$ and $g\from A \to Y$ in a Cartesian category, we write $\langle f, g \rangle \from A  \to X\times Y$ to denote the map induced by the universal property of the product.
\begin{definition}
  Let $F \from \RMod \to \RMod$ be a functor and $(A,f)$ and $(B,g)$ $F$-coalgebras.
\begin{enumerate}
\item
A \emph{state} of $(A, f)$ is an element $x \in A$.
\item
A \emph{bisimulation} $(p_1, p_2)$ between $(A,f)$ and $(B,g)$ is a span of coalgebra maps $A \xleftarrow{p_1} S \xrightarrow{p_2}B$ such that $\langle p_1, p_2 \rangle: S \to A\times B$ is a monomorphism.
\item
Two states  $x \in A$ and $y \in B$ are \emph{bisimilar} if there exists a bisimulation and an $s\in S$ such that $p_1(s) = x$ and $p_2(s)=y$.
\end{enumerate}
\end{definition}
In $\Set$ when $F$ preserves weak pullbacks, bisimulation is detected by maps into the final coalgebra in the sense that, $x,y \in A$ are bisimilar if and only if ${!_A}(x) = {!_A}(y)$ where $!_A \from A \to \nu F$ is the final coalgebra map \cite[Theorem 9.3]{Rutten00}.
This latter condition is called \emph{behavioural equivalence}.
Generally, bisimilarity and behavioural equivalence agree for weak pullback preserving functors~\cite{Staton2011-Bisim}.
\begin{theoremE}
\label{thm:usual}
If $F \from \RMod\to \RMod$ preserves weak pullbacks, and has a final coalgebra $\nu F$, then bisimilarity  and behavioural equivalence coincide on states in $F$-coalgebras.
\end{theoremE}
\begin{proofE}
Consider an $F$-coalgebra $(A, f)$ and let $!_A \from A\to \nu F$ be the final coalgebra map. By the definition of  behavioural equivalence, if one has two behaviourally equivalent states $x, y \in A$ then $x- y \in \ker{(!_A)}$. 
 Now, consider $S\coloneqq \setDef{(x,y)}{ x-y \in  \ker{(!_A)}  }$ i.e., the pullback $A\times_{\nu F}A$ of $!_A \from A \to \nu F$ along itself. This is equipped with the obvious projections $p_1, p_2 \from S \to A.$  We want to construct a map $p\from S\to FS$ such that the following diagram commutes: 
$$
\begin{tikzcd}
A \dar{f} & \arrow[l, "p_1", swap] S \arrow[d, dotted, "p"]  \rar{p_2}& A\dar{f}
\\
FA & \arrow[l, "Fp_1", swap] FS \rar{Fp_2} & FA.
\end{tikzcd}
$$
Reading off the diagram, we must have that $\langle f\circ p_1, f \circ p_2 \rangle =\langle Fp_1, Fp_2 \rangle  \circ p$. As $!_A$ is a coalgebra map, the map $\langle f\circ p_1, f \circ p_2 \rangle \from S \to FA \oplus FA$ factors through $FA\times_{F(\nu F)} FA$, seen as a submodule of $FA \oplus FA$, giving a diagram:
$$
\begin{tikzcd}
S \arrow[rr, "\langle f\circ p_1{,} f \circ p_2 \rangle"] \arrow[drr, "p", dotted, swap] && FA\times_{F(\nu F)} FA 
\\
\ && FS. \arrow[u, "\langle Fp_1{, }  Fp_2 \rangle", swap]
\end{tikzcd}
$$
The map $p$ exists because there is a section of the map $\langle Fp_1, Fp_2 \rangle$ in the above diagram.
This is because $F$ preserves weak pullbacks and $S$ is a pullback and so by the universal property of a weak pullback, there is a (not necessarily unique) map $FA\times_{F(\nu F)} FA \to FS$ induced by the limit map of the diagram defining $FA\times_{F(\nu F)} FA$.

Conversely if $S$ is a bisimulation with $ (x,y)\in S$, then $!_A(x) = !_S(x,y) = !_A(y)$ by finality of $\nu F$.
\end{proofE}
However, the weak pullback hypothesis may be unduly restrictive, but it can be improved for modules.
Suppose $\langle Fp_1{, }  Fp_2 \rangle  \from F\left(A\times_{\nu F}A \right)  \to FA\times_{F(\nu F)}FA$ is epi and that $[e] \in \operatorname{Ext}^1\left(FA\times_{F(\nu F)}FA, \ker \langle Fp_1{, }  Fp_2 \rangle \right)$ is the class of the following short exact sequence.
\begin{equation*}
0\to \ker \langle Fp_1{, }  Fp_2 \rangle \to F\left(A\times_{\nu F}A\right)\to FA\times_{F(\nu F)}FA \to 0
\end{equation*}
Then the relation $S$ carries a bisimulation iff $w^*[e] = 0$, where $w = \langle f p_1, f p_2\rangle$ and $w^*[e] \in \operatorname{Ext}^1\left(A\times_{\nu F}A, \ker \langle Fp_1{, }  Fp_2 \rangle \right)$.
Note that this is a direct refinement of the usual formulation of bisimulations in terms of relation lifting~\cite{Staton2011-Bisim}.
In particular this holds whenever $A\times_{\nu F}A$ is projective or $\ker \langle Fp_1{, }  Fp_2 \rangle$ is injective.
Thus, in $\RMod$ the axiom of choice is replaced by a cohomological obstruction.

For convenience, we phrase the discussion above as a proposition.
We see this proposition as providing clarity on some of the questions discussed at the end of \cite{bonchi12}, as for the case modules over a field (vector spaces) everything is projective and so the second condition is automatically fulfilled.
\begin{propositionE}
\label{prop:ext}
Let $F \from \RMod\to \RMod$ have a final coalgebra $\nu F$. If $(A, f)$ is a coalgebra such that:
\begin{enumerate}
\item
 the comparison map $\langle Fp_1{, }  Fp_2 \rangle \from F\left(A\times_{\nu F}A \right)  \to FA\times_{F(\nu F)}FA$ is epi;
 \item
  $w^*[e] = 0 \in \operatorname{Ext}^1(A\times_{\nu F}A , \ker \langle Fp_1{, }  Fp_2 \rangle )$, where $w = \langle f p_1, f p_2\rangle$.
\end{enumerate} 
Then bisimilarity  and behavioural equivalence are the same on states of $A$.
\end{propositionE}
\begin{proofE}
Fix the conventions established in the proof of Theorem \ref{thm:usual}.
The second condition gives us that 
\begin{equation*}
0 \to \ker \langle Fp_1{, }  Fp_2 \rangle \to \left(A\times_{\nu F}A\right) \times_B F(A\times_{\nu F}A ) \to A\times_{\nu F}A   \to 0
\end{equation*} 
splits, where $B= FA\times_{F(\nu F)}FA$.
Compose a section with the projection to $F(A\times_{\nu F}A )$, this gives the desired coalgebra structure.
So the desired lift of $w$ through $\langle Fp_1, Fp_2 \rangle$ exists if the conditions of the proposition are met.
\end{proofE}

\begin{example}
There are behaviourally equivalent states of affine functors that are not bisimilar. This is a behaviour that does not exist for the analogous functors in sets or vector spaces. For instance, consider the functor $F(X)= \Z/4\Z \oplus( \Z/2\Z \otimes X)$. We consider the coalgebra with carrier $Y = \Z/2\Z e \oplus \Z/ 2 \Z t$ and with structure map $\gamma = \langle o, f \rangle$, defined on the basis by:
\begin{equation*}
o(e) = 0, \qquad o(t) =2, \qquad f(e) = t, \qquad f(t) = 0 \, .
\end{equation*}

By computing the final coalgebra chain using that $(\Z/2\Z )\otimes (\Z/4\Z ) \cong \Z/2\Z$, we see that the final coalgebra of $F$ is obtained in $\omega$-steps as $(\Z/ 4\Z) \oplus (\Z/2 \Z)^{\omega}$. 
Let $\bar o \from Y \to \Z / 2 \Z$ be $o$ followed by reduction modulo $2$. 
In this coalgebra, we can compute the final coalgebra map applied to $e \in Y$ as:
\begin{equation*}
!_Y(e) = \left(o(e), \bar o(f(e)), \bar o(f^2(e)), \bar o(f^3(e)), \dots \right) = (0,0, 0, \dots).
\end{equation*}
The second stage here follows by noticing that $o(f(e)) = 2 \equiv 0 \pmod{2}$. The states $e$ and 0 are therefore behaviourally equivalent. 

However, no bisimulation $(S, \langle o', f' \rangle) \to Y \times Y$ contains the pair $(e, 0)$.
To see this, first observe that $2$ is in the annihilator of $Y \times Y$, and hence of $S$, so $\Z / 2 \Z \otimes S \cong S$.
It follows that $F(S) = \Z / 4 \Z \oplus S$.
Now, we compute $f'(e, 0) \in S$.
We must have $\pi_1(f'(e, 0)) = f(e) = t$ and $\pi_2(f'(e, 0)) = f(0) = 0$, so $f'(e, 0) = (t, 0) \in S$.
Unfolding one more step, one has that
$
2 = o(t) = o'(t, 0) = o(0) = 0.
$
This is not possible in $\Z / 4 \Z$ and therefore we have obtained the desired contradiction.
\end{example}

\paragraph*{The final coalgebra chain.}

Given the $\omega^{\mathrm{op}}$-chain (\ref{omegach}) in a locally finitely presentable category $\Cat{C}$, first let $\ell_n \colon V_\omega \to F^n 1$ ($n < \omega$) be the limit along with the cone morphisms. The universal property of the limit gives us a unique morphism $m \colon F V_\omega \to V_\omega$ such that for all $n \in \omega^{\mathrm{op}}$, we have
\begin{equation}
\begin{tikzcd}
F V_\omega \arrow[rr, "m_{\omega}"] \arrow[dr, "F \ell_n"'] & & V_\omega \arrow[dl, "\ell_{n+1}"] \\
& F^{n+1} 1 & 
\end{tikzcd}
\end{equation}
This yields a connecting morphism from $V_{\omega+1} = F V_\omega$ to $V_\omega$ allowing us to continue  transfinitely the $\omega^{op}$ chain (\ref{omegach}).

If $F$ preserves the limit $V_\omega$, then $m$ will be an isomorphism. We may therefore invert it giving a terminal coalgebra $m^{-1} \colon V_\omega \to F V_\omega$. Then we say that the final coalgebra \emph{converges in $\omega$ steps.}

In many cases the map $m_{\omega}$ fails to be an isomorphism. One needs to continue the iteration transfinitely, by iterating $F$ on the morphism $m \colon F V_\omega \to V_\omega$ , obtaining the $\omega^{\mathrm{op}}$-chain
\begin{equation}
V_\omega \xleftarrow{m} V_{\omega+1} \xleftarrow{Fm} V_{\omega+2} \xleftarrow{F F m} \cdots .
\end{equation}
Its limit is denoted by $V_{\omega \cdot 2} = \lim_{n<\omega} V_{\omega+n}.$
The question becomes when $F$ preserves this limit. If it does, by the same argument as before, we can obtain the coalgebra structure on it by inverting the isomorphism. 

Ad\'amek, Milius and Moss \cite{AdamekEA25}, following Worrell \cite{Worrell99,Worrell05},  give sufficient conditions on $F$ and $\Cat{C}$. To describe this we shall need the following omnibus definition. Note that in $\RMod$ the situation simplifies dramatically as every intersection is nonempty and (5) will be the only condition that matters in the remaining part of the paper.

\begin{definition}
  In a category $\C$ we say that:
\begin{enumerate}
\item
An initial object $0$ is \emph{strict} if every morphism with codomain $0$ is an
isomorphism.
\item
 A monomorphism $A \hookrightarrow B$ is \emph{empty} if its domain is a strict initial object; it is
\emph{nonempty} if it is not empty.
\item
An intersection, i.e., a wide pullback of monomorphisms, is \emph{empty} if its domain is a strict
initial object.
\item
An endofunctor $F \from \C \to \C$ preserves nonempty intersections if $F$ takes a nonempty
intersection to a nonempty intersection.
\item
A functor $F \from \C \to \C$ preserves \emph{countable directed intersections of monomorphisms} if, for every $\omega^{op}$ chain $V_i$ where the connecting maps are monomorphisms, the canonical comparison map $ F(\lim_{i\in \mathbb N} V_i) \to \lim_{i\in \mathbb N} F\left(V_i\right)$ is an isomorphism.
\end{enumerate}
\end{definition}
It is noted in \cite{AdamekEA25} that preserving binary intersections implies monomorphism preservation. In $\RMod$, the last assumption also implies monomorphism preservation.
\begin{propositionE}
\label{prop:monopres}
If a functor $F \from \RMod \to \RMod$ preserves countable directed intersections of monomorphisms, then it preserves monomorphisms.
\end{propositionE}
\begin{proofE}
We prove this by first establishing the following categorical lemma.
\begin{lemma}
\label{lem:ready}
Let  $m \from A \to X$ be a monomorphism. Suppose that there exists a chain of inclusions $D_0\hookleftarrow D_1 \hookleftarrow \cdots$ with limit cone $(A, \ell_n)$, and such that each $\ell_n$ factors through $m$. Then any functor that preserves this limit sends $m$  to a mono.
\end{lemma}
\begin{proof}
Suppose that $Fm\circ u = Fm \circ v.$ Postcomposing with the maps that give the limit cone, we obtain that $F(\ell_n) \circ u = F(\ell_n) \circ v$. Since $F(\ell_n)$ is assumed to be a limit cone, we know it is jointly monic, whence $u = v.$
\end{proof}
So now to prove the proposition, it remains to find such a chain.
To that end, let  $m \from A \to X$ be a monomorphism. 
We prove the proposition by exhibiting a chain as in Lemma \ref{lem:ready}. 
We define a module $M$ by taking $\bigoplus_{i=0}^\infty X$ and quotienting by  $\setDef{a_i - a_j}{ a \in A},$ where $a_i$ denotes the component of $a_i$ in the $i^{th}$ copy of $X$. 
In other words, we coequalise over possible morphisms $m_i \from A \to X \xrightarrow{p_i} \bigoplus_{i=0}^\infty X$.
There is an obvious way to  filter $M$, take 
\begin{equation*}
D_0= M \qquad D_n = \operatorname{Im}\left(\bigoplus_{i=n}^\infty X\longrightarrow M \right)
\end{equation*}
The common map $A \to M$ is monic, since its composite with the map $M \to X$ induced by summing the components is $m$.
Moreover, $M / A \cong \bigoplus_{i=0}^\infty (X / A)$, and the images of the $D_n$ are the corresponding tails, whose intersection is zero.
The limit is clearly $A$, and the map $X \xrightarrow{p_n}  \bigoplus_{i=0}^\infty X$ factors $p_n' \from X \to D_n$  so the limit map factors as $\ell_n = p_n' \circ m$
We apply Lemma \ref{lem:ready} to the above to obtain our result.
\end{proofE}
\begin{example}
The initial object in $\RMod$ is strict if and only if $R$ is the trivial ring i.e., the ring with one element. This is because $\RMod$ has a zero object, which implies that it admits a morphism from every object, and such an object can only be strict if it is the only object in the category. This only happens when $R$ is the trivial ring. It follows that all intersections in $\RMod$ are nonempty. On the other hand, unlike in vector spaces, not all finitary functors in $\RMod$ preserve binary intersections.
\end{example}
\begin{definition}
A locally finitely presentable category $\Cat{C}$ is said to satisfy the descending chain condition (DCC) and be a \emph{DCC-category} if every descending chain of subobjects or strong quotients of finitely presentable objects stabilises.
\end{definition}
\begin{example}
\label{Ex:artdcc}
The category $\RMod$  is DCC when $R$ is Artinian. This is because a finitely generated $M$ over a Noetherian ring $R$ has finite length if and only if $R/\Ann M$ is Artinian \cite[Corollary 2.17]{eisenbud95}, and it is easy to see that quotients of Artinian rings are Artinian by the correspondence between ideals of $R$ containing $I$ and ideals of $R/I$.
\end{example}
Ad\'amek, Milius and Moss prove the following result~\cite{AdamekEA25}.
\begin{theorem}
Let $\Cat{C}$ be a DCC-category. Every finitary endofunctor $F \from  \Cat{C} \to \Cat{C}$ which
preserves nonempty binary intersections has a final coalgebra obtained in  $\omega \cdot 2$  steps.
\end{theorem}
This instantiates to the class of Artinian rings.
\begin{corollaryE}
\label{cor:artin}
If $R$ is an Artinian ring, then all finitary endofunctors on $\RMod$ that preserve binary intersections converge in at most $\omega \cdot 2$ steps.
\end{corollaryE}
\begin{proofE}
We already noted that $\RMod$ is an lfp category. All binary intersections are nonempty in $\RMod$. So, to prove the corollary, we must show that it satisfies the DCC condition. We showed this in Example \ref{Ex:artdcc}.
\end{proofE}
\begin{remark}
As all fields are Artinian, the corollary extends the well-known fact that all finitary functors on vector spaces converge in at most $\omega \cdot 2$ steps.
\end{remark}
The best known non-Artinian ring is $\Z.$
 In the next section, we shall show that the convergence theory fails to extend to $\ZMod$.

\section{$\omega\cdot 2$-Steps is Not Enough}
\label{sec:counterexample}
\pratendSetLocal{category=counterexample}
Fix a prime number $p$. We exhibit a finitary functor~$U \colon\ZMod\to\ZMod$ preserving binary intersections whose final sequence does not stabilise after $\omega\cdot 2$ steps.  The category $\ZMod$, contains a full subcategory, the torsion modules, that is generated by the finitely presentable objects that satisfy the DCC-condition. This subcategory, as a subcategory, is not closed under $\omega^{op}$-limits. We first construct a functor $F_p$ that has a completely torsion final coalgebra sequence at every finite stage but which has elements of infinite order in the limit. There is a second category in the picture here, that of $\Q$-vector spaces. Tensoring by $\Q$ kills every torsion group, so if we apply the functor $F_p(-)\oplus (W \otimes (\Q\otimes -))$, the second component disappears on every finite stage, but suddenly becomes nonzero once the non-torsion appears. This restarts the building process and prevents termination at $\omega\cdot 2$. The deeper picture  motivating this construction will become clearer in Section~\ref{sec:omega3} when we discuss the Gabriel filtration.

 \paragraph*{A method for extending a group.}
 In this section, we explain a way of constructing non-split extensions. It is more intuitive to describe the following construction in terms of groups than modules, and the module-theoretic rephrasing can be easily worked out using the dictionary of Example \ref{abeliangroups}. The idea is to merge two groups together using an operation similar to \emph{carrying} in primary school arithmetic; for a fun paper that makes this connection more precise, see \cite{isaksen02}.

As a first step towards this goal, let $(G, \star, e)$ be an abelian group,  $g \in G$ and $p$ a prime number.
The \emph{$p$-extension of $G$ at $g$} is obtained by equipping the set $G_g := G \times\Z/ p\Z$ with the operation
\begin{equation*}
\bar\star\colon G_g\times G_g\to G_g,
\quad
(g_1, i) \bar\star (g_2, j) 
 =
\begin{cases}
        (g_1\star g_2, i+j), &\text{if}\quad i+j < p 
        \\ 
        (g_1\star g_2 \star g, i+j \bmod p ), &\text{if}\quad i+j \geq p
\end{cases}
\end{equation*}
Thus, intuitively, the operation $\bar\star$ marks overflows in~$\Zm{p}$ by scaling along~$\star$ by~$g$ in the first component. 
\begin{propositionE}
\label{prop:extension}
$G_g$ carries the structure of an abelian group with the operation~$\bar\star$ and unit $(e,0)$.
\end{propositionE}
\begin{proofE}
First, the operation on the set is clearly commutative, as $\star$ is commutative in $G$ and addition is commutative in $\mathbb{Z}$.
We have that $(e,0)\bar \star(h, i) = (e\star h, i) = (h,i)$ since $0+i = i \leq p-1$, so the identity element is given by $(e, 0)$.
    
Inverses are given by the formula
\begin{equation*}
(h,i)^{-1} = 
\begin{cases}
(h^{-1}, 0) &\mbox{ if } i =0
\\
(h^{-1} \star g^{-1}, p-i) &\mbox{ if } i \neq 0
\end{cases}
\end{equation*}
where $h^{-1}$ denotes the inverse of $h$ in $G$.

Finally, we verify associativity. Let $x=(h_1, i)$, $y=(h_2, j)$, and $z=(h_3, k)$. 
The second component is simply addition modulo $p$, which is associative. 
The first component consists of the term $h_1 \star h_2 \star h_3$ multiplied by $g$ whenever a carry occurs in the indices. 
The number of times we add $g$ to the first component is equal to $\lfloor (i+j+k)/p \rfloor$. 
Since this value depends only on the total sum and not on the grouping of terms, the operation is associative.
\end{proofE}

\begin{example}
We have a short exact sequence
\begin{equation*}
0 \to G \to G_g \to \Z/p\Z \to 0
\end{equation*}
For $g=e$, this sequence is split, so that one has  $G_e \cong G\oplus \Z/ p\Z$. For any element $g$ in $\Z/p^n \Z$ that is not divisible by $p$, we obtain $\left(\Z/p^n \Z \right)_g = \Z/p^{n+1} \Z.$
\end{example}

 \paragraph*{Group extensions, functorially.}

The previous construction depended on making a choice of $g \in G,$ and so is not functorial in $G$. To extract a functor from this construction we take all such choices of $g$ simultaneously and add them using the direct sum.
\begin{definition}
\label{def:p-ification_functor}
The \emph{group extension functor} $F_p: \ZMod \to \ZMod$ is defined on objects by
\begin{equation*}
    F_p(G) = \bigoplus_{g\in G} G_g.
\end{equation*}
On morphisms $f\from G \to H$, it is given by $F_p(f): \bigoplus_{g\in G} G_g \to \bigoplus_{h\in H} H_h$ which maps the summand $G_g$ to $H_{f(g)}$ via the assignment $(a, i) \mapsto (f(a), i).$
\end{definition}

We now must verify that the group extension functor has the properties needed by a counterexample.
\begin{propositionE}
The functor $F_p$ is finitary and preserves binary intersections.
\end{propositionE}
\begin{proofE}
We check the conditions independently.
\begin{lemma}
The functor $F_p$ is finitary.
\end{lemma}
\begin{proof}
Let $\mathcal D \from \mathcal{I} \to \ZMod$ be a filtered diagram with colimit $G = \colim_{i \in \mathcal{I}} \mathcal D^i$.
There is a canonical map $\psi:\colim_{i\in\mathcal I} F_p(\mathcal D^i)\to F_p(G)$.
To prove the result, we must show that this is bijective and thus an isomorphism since we are working in abelian groups (which is an balanced category). 
We use the concrete description of filtered colimits given in Proposition \ref{prop:AbFC} to do this.

Let $x \in F_p(G)$. It has finite support so we may write it as the finite sum $x =\sum_{r=1}^m x_{g_r}$, with  each $x_{g_r}=(a_r,s_r) \in G_{g_r}$.
By Proposition \ref{prop:AbFC}, all the elements $g_r$ and $a_r$ sit in some common element of the directed colimit $\mathcal D^i$.
Each component $(\mathcal D^i)_h$ is set-theoretically $\mathcal D^i \times \{0, \dots, p-1\}$, so these lifts give a preimage of $x$ in $F_p(\mathcal D^i)$.
Thus $x$ is in the image of $F_p(\mathcal D^i)$, and consequently in the image of the colimit.
So $\psi$ is surjective.

Now suppose that there exists $y \in \colim_{i \in \mathcal I} F_p(\mathcal D^i)$ such that $\psi(y) = 0$.
Proposition \ref{prop:AbFC} gives an index $i$ such that $y$ has a representative $y_i' \in F_p(\mathcal D^i)$.
Since $F_p(\mathcal D^i) = \bigoplus_{h\in\mathcal D^i}(\mathcal D^i)_h$, it follows that $y_i'$ has finite support.
These terms sum to zero in $F_p(G)$ as $\psi(y)=0$.
Since only finitely many summand indices are involved, we may pass to a common later stage where all indices with the same image in $G$ become equal.
After adding the terms with equal indices, each remaining component maps to zero separately.
Its second coordinate is therefore zero, and its first coordinate becomes zero at some later stage.
As there are only finitely many components, there is a transition morphism $i\to j$ such that the image of $y_i$ vanishes in $F_p(\mathcal D^j)$.
So $y=0$ as required. So $\psi$ is injective.

Therefore $F_p$ is finitary.
\end{proof}
It is almost trivial to show that the group extension functor preserves monomorphisms because it does so on each component.
\begin{lemma}
\label{lem:Fp-monos}
The functor $F_p$ preserves monomorphisms.
\end{lemma}
\begin{proof}
Let  $f \from G \to H$ be monic. Then $F_p(f)$ is the morphism induced by the injective index map $f \from G \to H$ together with the components $G_g \to H_{f(g)}$, $(a,i) \mapsto (f(a), i)$, each of which is monic because $f$ is. Direct sums of monos are monos and the conclusion follows.
\end{proof}
Hence $F_p$ preserves binary intersections.
\begin{lemma}
The functor $F_p$ preserves binary intersections.
\end{lemma}
\begin{proof}
From Lemma \ref{lem:Fp-monos}, we have that $F_p$ preserves monomorphisms. Given subgroups $A$ and $B$ of $G$, as $F_p$ preserves monomorphisms, we can regard $F_p(A)$ and $F_p(B)$ as subobjects of $F_p(G)$.	

    Suppose $x \in F_p(G)$ is in the intersection $F_p(A) \cap F_p(B)$. Then, the support of $x$ is in the intersection $A\cap B.$ Moreover,  for every $g \in \operatorname{supp}(x)$, the component $x_g$ must belong to $A_g \cap B_g$.  Observe that 
     $$
    A_g \cap B_g = (A \times \{0, \dots, p-1\}) \cap (B \times \{0, \dots, p-1\}) = (A \cap B)_g.
    $$
    Therefore,
    $$
    F_p(A) \cap F_p(B) = \bigoplus_{g \in A \cap B} (A \cap B)_g = F_p(A \cap B),
    $$
    as required.
\end{proof}
\end{proofE}
\paragraph*{The $\omega$-step of the final coalgebra chain for the group extension functor.}
We now consider the final coalgebra construction for $F_p$. The point of this section is to prove that $V_\omega$ contains an element of infinite order. To aid the reader's intuition, we shall start with an example and spend the remainder of this subsection formalising it.
\begin{example}
For the first step, $V_1 = F_p(0) = 0_0 = \Z/p\Z$. The connecting map $V_1 \to V_0$ is, as usual, the zero morphism.
The second stage is $V_2 = \bigoplus_{g \in \Z/p\Z} (\Z/p\Z)_g$. The connecting map sends a given pair $(a, j)$ to $j$. In particular $(\Z/p\Z)_1 = \Z/p^2\Z.$
So, the chain of objects contains a sub-sequence of cyclic groups of increasing order:
\begin{equation*}
\begin{tikzcd}
\cdots \rar& \Z/ p^3\Z  \rar &   \Z/ p^2\Z \rar & \Z/ p\Z \rar  & 0 
 \end{tikzcd}
\end{equation*}
The limit of this chain is well-known to be the $p$-adic integers $\pZ$ \cite[Section 7.1]{eisenbud95}, which is a torsion free group.
The intuition that the reader should keep in mind is that, at each stage the order of the torsion increases, and so, in the limit, one obtains an element of infinite order, yielding a group that is no longer completely torsion.
\end{example}
We now formalise the intuition from the previous example to study the first $\omega$ steps of the final coalgebra chain.
 First, note that each $V_n$ is a finite group.
\begin{lemmaE}
\label{lem:stages-torsion}
Each $V_n$ is a finite and therefore torsion group.
\end{lemmaE}
\begin{proofE}
By induction on $n$: $V_0 = 0$, and if $G$ is a finite, torsion group then each $G_g $ is a finite torsion group by direct inspection, and hence so is $F_p(G) = \bigoplus_{g\in G} G_g$.
\end{proofE}
Next, we construct a very explicit coherent limit sequence where the order of the element tends towards infinity as $n$ gets larger.
\begin{lemmaE}
\label{lem:tracked}
We define elements $z_1 \coloneqq (0_{V_0},1) \in (V_0)_0 = V_1$ and, by iteration, $z_{n+1} \coloneqq (0_{V_n},1) \in (V_n)_{z_n} \subseteq F_p(V_n) = V_{n+1}$.
Then for every $n\ge 1$, 
(i) $z_n$ has order $p^n$ and 
(ii) $m_n(z_{n+1}) = z_n$.
\end{lemmaE}
\begin{proofE}
\begin{enumerate}
\item
In $G_g$, it is easily seen that $k\cdot(0,1) = (q\cdot g, r)$ for $k = qp + r$ with $0\le r<p$.
By induction, it follows that if $z_n$ has order $p^n$, then $p^{n+1}\cdot(0,1) = (p^n z_n, 0) = 0$.  
Further,  observe that $p^{n}\cdot(0,1) = (p^{n-1} z_n, 0) \neq 0$ in $(V_n)_{z_n}.$  
\item
We proceed by induction. For the base case of  $n=1$ notice that $m_1 = F_p(m_0)$ maps the summand $(V_1)_{z_1}$ to $(V_0)_{m_0(z_1)} = (V_0)_0 = V_1$ and so $m_1(z_2)=(0,1)= z_1$. 
Assuming the inductive hypothesis that  $m_n(z_{n+1}) = z_n$, we have $m_{n+1}=F_p(m_n)$ maps $(V_{n+1})_{z_{n+1}}$ to $(V_n)_{m_n(z_{n+1})}=(V_n)_{z_n}$ by $(a,i)\mapsto (m_n(a), i)$. We conclude that $m_{n+1}(z_{n+2})= (0,1) =z_{n+1}$ as desired.
\end{enumerate}
\end{proofE}
We deduce from this that the limit $V_\omega$ has a non-torsion subgroup, generated by the element $z$ that we just constructed.
\begin{corollaryE}
\label{cor:padic}
The tuple $z = (z_n)_{n<\omega}$ is an element of the limit $V_\omega$ of infinite order. In particular $V_\omega$ is not torsion.
\end{corollaryE}
\begin{proofE}
From Lemma \ref{lem:tracked}(2), it follows that the tuple is an element of the limit.
It is obvious that it is non-torsion: suppose that we have $kz = 0$ for some $k \ge 1$. 
This would imply that $k z_n = 0$ for all $n$, so we then have $p^n \mid k$ for all $n$ by Lemma \ref{lem:tracked}(1),  which is a contradiction.
The subgroups $\langle z_n\rangle \cong \Z/p^n\Z$ are mapped onto each other by the connecting maps, and an inverse limit of subgroups is a subgroup of the inverse limit,  whence we obtain the $p$-adic integers $\pZ = \lim_n \Z/p^n\Z \hookrightarrow V_\omega$.
\end{proofE}
We turn next to constructing the functor that will dominate the second block of $\omega$ steps by detecting the appearance of non-torsion.
\begin{definition}
  Fix a $\Q$-vector space $W$ with a countable basis $\{v_i\}_{i \in \mathbb N_0}$.
  We define the functor $T\from \ZMod \to \ZMod$ by composing the functor $M \mapsto \Q \tens_{\Z} M$ with the tensor product (over $\Q$) with the vector space $W$:
\begin{equation*}
T(M) = W \otimes_{\Q}\left( \Q \otimes_{\Z}  M\right)
\end{equation*}
\end{definition}
Crucially, $T(G)=0$ for any torsion group $G$.
The infinite basis of $W$ lets us construct a coherent sequence using a new basis vector at each successive stage after $\omega$ in the final coalgebra chain.
The resulting limit element can be viewed as an infinite sum, but any preimage under $m_{\omega\cdot 2}$ would have to have a $T$-component involving only finitely many basis vectors, which will give us our proof of nonconvergence at $\omega \cdot 2$.

The basis induces an exhaustive filtration $\mathcal F^jW$ on $W$. This consists of all elements that can be expressed as finite linear combinations of basis elements $\{v_i\}_{i \in \mathbb N_0,  i \le j}.$ We write $|v|$ for the unique minimal integer such that $v \in \mathcal F^{|v|}W$.

\begin{propositionE}
For any $M \in \ZMod$, the submodules
\begin{equation*}
\mathcal F^j T(M) \coloneqq \mathcal F^j W \otimes_{\Q} \left(\Q \otimes M
\right) \subseteq W \otimes_{\Q} \left(\Q \otimes M\right) = T(M)
\end{equation*}
form an exhaustive filtration of $T(M)$.
\end{propositionE}
\begin{proofE}
Each $\mathcal F^j W$ is a direct summand of $W$ as a $\Q$-vector space, so $\mathcal F^j T(M)$ is a submodule of $T(M)$.
Since $W = \bigcup_j \mathcal F^j W$ and tensor products commute with directed colimits, we have that $T(M) = \bigcup_j \mathcal F^j T(M)$.
\end{proofE}
We establish the categorical properties of $T$ required for the main theorem.
\begin{propositionE}
The functor $T$ is finitary and preserves binary intersections.
\end{propositionE}
\begin{proofE}
   First we show that $T$ is finitary. The tensor product $-\tens_{\Z} \Q$ is a left adjoint to the hom-functor and thus preserves colimits. Since filtered colimits commute with finite tensor products and direct sums, the whole functor $T$ is finitary.

It is easy to see that binary intersections are a type of finite limit.
As a $\Z$-module,  $\Q$ is torsion-free and over a principal ideal domain like $\Z$, all torsion-free modules are flat \cite{weibel94}. 
Therefore, the functor $-\otimes_{\Z} \Q$ is exact and preserves all finite limits, including binary intersections. Tensoring with $W$ is exact so it also preserves all finite limits.
   
\end{proofE}

\paragraph{The counterexample functor.}
We are finally in a position to introduce the functor that will not converge in $\omega \cdot 2$ steps.
$$
U(G) \coloneqq  T(G)\oplus F_p(G)
$$
This is a finitary functor and preserves binary intersections because both $T$ and $F_p$ do and so does the direct sum.

The following easy lemma shows that we have already computed the first $\omega$-steps of the final coalgebra sequence attached to $U$.
\begin{lemmaE}
The final coalgebra sequence of $U$ coincides with the final coalgebra sequence of $F_p$ for the first $\omega$ steps.
\end{lemmaE}
\begin{proofE}
The functor $T$ vanishes on torsion groups. All of the groups appearing in the first $\omega$ stages of the final coalgebra sequence for $U$ are torsion. The conclusion follows.
\end{proofE}
\begin{remark}
In $\Set$ and other DCC-categories, the connecting map $FV_\omega \to V_\omega$ is always a monomorphism and hence the second stage of $\omega$-steps can be thought of as a process of refining the algebra $V_{\omega}$.
The map $m_{\omega} \from UV_{\omega} \to V_{\omega}$ is not a monomorphism, meaning this pattern breaks in $\ZMod.$
However, it is still possible to see $m_{\omega}$ as a generalised monomorphism: it just has torsion-free kernel instead of zero kernel.
We shall exploit this fact in Section \ref{sec:omega3} to prove our novel convergence result.
\end{remark}
We shall not need the full structure of $V_{\omega +k}$, finding a well-behaved submodule will suffice. 
The following proposition isolates a convenient subsequence of the second block of $\omega$-steps. 
The proofs below are delicate chiefly due to the fact that $F_p$ is nonadditive, so we reason our way around this using injections instead.
\begin{propositionE}
\label{splitting}
For each $k\in \mathbb N$, there is an injection 
\begin{equation*}
\phi_k \colon \bigoplus_{i=0}^{k} T^i F_p^{k-i } V_{\omega} \hookrightarrow V_{\omega +k} 
\end{equation*}
where $T^0 = F_p^0 = \id.$ Moreover, these injections are compatible with the connecting maps in the sense that the following diagram commutes:
\begin{equation*}
\begin{tikzcd}[ampersand replacement=\&]
 \bigoplus_{i=0}^{k+1} T^i F_p^{k+1-i } V_\omega \dar{\phi_{k+1}}  \arrow[rrr, "b_k"] \&\&\& \bigoplus_{i=0}^{k} T^i F_p^{k-i } V_\omega \dar{\phi_k}
\\
V_{\omega +k+1} \arrow[rrr, "m_{\omega + k}"] \&\&\& V_{\omega +k}.
\end{tikzcd}
\end{equation*}
with $b_k$ defined to be $\bigoplus_{i=0}^{k} T^i F_p^{k-i} (m'_{\omega})$ on the summand $\bigoplus_{i=0}^{k} T^i F_p^{k+1-i } V_\omega$  and identically $0$ on the summand $T^{k+1}V_{\omega}$.
Here $m'_{\omega}$ denotes the restriction  $m_{\omega}|_{F_p V_{\omega} }$.
\end{propositionE}
\begin{proofE}
First, the connecting map $m_{\omega}|_{T V_{\omega}} \from TV_{\omega} \to V_{\omega}$ is the zero map. To observe this, note that, given the  projection $\ell_n: V_{\omega} \to V_n$, we have that $T(\ell_n) = 0$, since the codomain $T(V_n ) = 0.$ As the connecting map $m_\omega$ is defined by the universal property that $T(\ell_n)(x) = \ell_{n+1}m_{\omega}(x)$; it follows that  $m_{\omega}|_{T V_{\omega}} = 0 $. It follows that the non-zero part is the restriction $m_{\omega}'.$

We then prove the claim of the statement by induction.
\begin{enumerate}
\item
 For the base case, the inclusion $\phi_0 \from F_p^0 V_{\omega} \to  V_{\omega}$ is the identity map which is automatically mono. 
\item
  Assume by the inductive hypothesis now that we have a monomorphism  $\phi_k \from  \bigoplus_{i=0}^{k} T^i F_p^{k-i } V_\omega \to  V_{\omega+k}$.
  Recall that $V_{\omega+k +1} = U(V_{\omega+k}) = T(V_{\omega+k}) \oplus F_p(V_{\omega+k}).$ The map $\phi_{k+1}$ is defined to be the composition
\begin{equation*}
 \bigoplus_{i=0}^{k} T^{i+1} F_p^{k-i } V_\omega \oplus F_p^{k+1}(V_{\omega}) \xrightarrow{T\phi_k \oplus  F_p (\psi_k)} T(V_{\omega+k}) \oplus F_p(V_{\omega+k}) = V_{\omega+k+1}.
\end{equation*}
\end{enumerate}
We have that $F_p (\psi_k) \from  F_p^{k+1}(V_{\omega}) \to F_p(V_{\omega+k})$ is a monomorphism. This is because $\psi_k$ is a mono by Lemma \ref{babysplitting} and $F_p$ preserves monos. Since $T$ preserves monomorphisms and the direct sum of monos is also a mono it follows that $\phi_{k+1}$ is a monomorphism as desired.

Finally, we observe that the diagram 
\begin{equation*}
\begin{tikzcd}
 \bigoplus_{i=0}^{k+1} T^i F_p^{k+1-i } V_\omega \dar{\phi_{k+1}}  \arrow[rrr, "b_k"] &&& \bigoplus_{i=0}^{k} T^i F_p^{k-i } V_\omega \dar{\phi_k}
\\
V_{\omega +k+1} \arrow[rrr, "m_{\omega + k}"] &&& V_{\omega +k}
\end{tikzcd}
\end{equation*}
commutes by induction, as we iteratively apply $U$ to diagrams starting with
\begin{equation*}
\begin{tikzcd}
  F_p V_\omega \dar{\psi_{1}}  \arrow[rrr, "m'_{\omega}"] &&&  V_\omega \dar{\psi_0}
\\
 T(V_\omega)  \oplus F_p(V_{\omega}) = V_{\omega +1} \arrow[rrr, "m_{\omega}"] &&& V_{\omega}.
\end{tikzcd}
\end{equation*}
and extending by the zero map.
 \end{proofE}
 \begin{textAtEnd}[category=counterexample]
The proof of this result is an inductive argument where it is helpful to first prove the following simple case, where we consider only the leading term in the direct sum. The general case above adds the $T$-summands.

\begin{lemma}
\label{babysplitting}
For each $k\in \mathbb N$, there is a monomorphism 
\begin{equation*}
\psi_k \colon  F_p^k V_{\omega} \hookrightarrow V_{\omega +k}.
\end{equation*}
Moreover, these monomorphisms are compatible with the connecting maps, by which we mean that the following diagram commutes
$$
\begin{tikzcd}[ampersand replacement=\&]
F_p^{k+1} V_\omega \dar{\psi_{k+1}}  \arrow[r, "F_p^k m'_{\omega}"] \& F_p^{k} V_\omega \dar{\psi_k} \\
V_{\omega +k+1} \arrow[r, "m_{\omega+k}"] \& V_{\omega +k}.
\end{tikzcd}
$$
where $m'_{\omega}$ denotes the restriction $m_{\omega}|_{F_pV_{\omega}}$.
\end{lemma}

\begin{proof}
We then prove the claim of the statement by induction. 
The inclusion $\psi_0 \from F_p^0 V_{\omega} \to  V_{\omega}$ is the identity map which is always mono. 
By the inductive hypothesis, one may assume that we have a monomorphism  $\psi_k \from  F_p^k V_\omega \to  V_{\omega+k}.$ 
Recall that $V_{\omega+k +1} = U(V_{\omega+k}) = T(V_{\omega+k}) \oplus F_p(V_{\omega+k}).$ 
By definition, the map $\psi_{k+1}$ is the composition:
\begin{equation*}
F_p^{k+1}V_\omega \xrightarrow{F_p\psi_k} F_pV_{\omega +k } \xrightarrow{i} T(V_{\omega+k}) \oplus F_p(V_{\omega+k}) = V_{\omega+k+1}.
\end{equation*}
Since $F_p$ preserves monomorphisms and $\iota_{V_{\omega+k}} \from F_p(V_{\omega+k}) \to T(V_{\omega+k}) \oplus F_p(V_{\omega+k})$ is a summand inclusion, $\psi_{k+1} \coloneqq \iota_{V_{\omega+k}} \circ F_p(\psi_k)$ is a monomorphism. 
For the compatibility, note that $\iota \from F_p \Rightarrow T \oplus F_p = U$ is a natural transformation. 
We prove this by induction, note that the base square commutes because $m_\omega \circ \iota_{V_\omega} = m'_\omega$ by the definition of $m'_\omega$. 
For the step, using $m_{\omega+k+1} = U(m_{\omega+k})$, naturality of $\iota$ at $m_{\omega+k}$, and the inductive hypothesis $m_{\omega+k} \circ \psi_{k+1} = \psi_k \circ F_p^k m'_\omega$:
\begin{align*}
m_{\omega+k+1} \circ \psi_{k+2}
&= U(m_{\omega+k}) \circ \iota_{V_{\omega+k+1}} \circ F_p(\psi_{k+1})
= \iota_{V_{\omega+k}} \circ F_p(m_{\omega+k}) \circ F_p(\psi_{k+1})\\
&= \iota_{V_{\omega+k}} \circ F_p\bigl(m_{\omega+k} \circ \psi_{k+1}\bigr)
= \iota_{V_{\omega+k}} \circ F_p(\psi_k) \circ F_p^{k+1} m'_\omega
\\
&= \psi_{k+1} \circ F_p^{k+1} m'_\omega.
\end{align*}
\end{proof}
 \end{textAtEnd}
Finally, we use the previous description to form a coherent subsequence of the second $\omega$-chain, thus yielding an element of $V_{\omega \cdot 2}$
\begin{propositionE}
\label{prop:limit_descrip}
Let $z\in V_{\omega}$ be the element constructed in Corollary~\ref{cor:padic}. There is a sequence of elements $z^{(k)} \in F_p^k V_{\omega}$ such that:
\begin{enumerate}
\item
we have $F_p^{k-1}(m'_{\omega})(z^{(k)}) = z^{(k-1)}$ for every $k\geq 1$.
\item
we have that $z^{(k)}$ generates a submodule of $F_p^k V_{\omega}$ isomorphic to $\Z.$
\end{enumerate}
\end{propositionE}
\begin{proofE}
The construction of the sequence is by induction. First, set $z^{(0)} = z.$ Assume that $z^{(l)} \in F_p^l V_\omega$ is given. Then we define $(0,1) \in (F_p^{l} V_{\omega})_{z^{(l)}}$ to be $z^{(l+1)}.$ 

Condition (2) is now trivial as $(0,1)$ clearly has infinite order. 

Condition (1) is somewhat less trivial.
We have $m'_{\omega}(z^{(1)})=z$, since its projection to $V_{n+1}$ is $(0,1)=z_{n+1}$ for every $n\geq 0$. Unravelling this, we have the following diagram.
$$
\begin{tikzcd}
\left(V_{\omega}\right)_z  \dar{F_p(\ell_n )} \arrow[r, hook]& F_p(V_{\omega}) \dar{F_p(\ell_n)}\rar{m'_{\omega}} & V_\omega \arrow[dl, "\ell_{n+1}"]
\\
\left(V_n\right)_{\ell_n(z)} \arrow[r, hook]& V_{n+1}
\end{tikzcd}
$$
On the other hand, the connecting map $TV_{\omega} \to V_{\omega}$ is identically 0 as $T(\ell_i) = 0$ for all the limit cone maps.

For $l\geq 1$, assume that $F_p^{l-1}(m'_{\omega})(z^{(l)})=z^{(l-1)}$. By the definition of $F_p$ on morphisms, $F_p^l(m'_{\omega})$ sends $(0,1)\in (F_p^lV_{\omega})_{z^{(l)}}$ to $(0,1)\in (F_p^{l-1}V_{\omega})_{z^{(l-1)}}$. These elements are $z^{(l+1)}$ and $z^{(l)}$, proving the induction step.
\end{proofE}
We can finally prove our main result.
\begin{theorem}
The final coalgebra sequence of $U$ does not converge in $\omega \cdot 2$ steps.
\end{theorem}
\begin{proof}
We use Proposition \ref{splitting} to give an element in $V_{\omega \cdot 2 }$ that is not in the image of the connecting map. First, we consider the element $z \in V_{\omega}$, constructed in Corollary~\ref{cor:padic}. We use the basis $(v_i)$ of $W$ fixed above. Set $w_0 = 0$ and then for each $V_{\omega + k}$, where $k>0$, we fix the element
\begin{equation*}
w_k  \coloneqq \sum_{i = 1}^k v_i \otimes v_{i-1} \cdots \otimes v_1 \otimes \left( 1_{\Q} \otimes z^{(k-i)}\right).
\end{equation*}
We regard this sum as an element of $V_{\omega + k}$ via the monomorphism $\phi_k$ of Proposition \ref{splitting}.
Here $z^{(k-i)} \in F_p^{k-i} V_{\omega}$ is the element given by Proposition \ref{prop:limit_descrip}. By Proposition~\ref{splitting} and
Proposition~\ref{prop:limit_descrip}(1), the $(\cdots, w_3, w_2, w_1, w_0)$ is coherent and determines some $v \in V_{\omega \cdot 2 }.$ This element can be thought of as a sum with infinitely many terms.

It is now easy to show by contradiction that $v$ is not in the image of the connecting map $m_{\omega \cdot 2 }$.
Suppose that there was a $u\in V_{\omega\cdot 2 +1}$ such that $m_{\omega \cdot 2}(u) = v.$ 
Let $u= (u_T, u_F)$. where $u_T\in T(V_{\omega\cdot 2})$ and $u_F \in F_p(V_{\omega \cdot 2}).$ 
As the filtration of $T(V_{\omega \cdot 2})$ is exhaustive, we can choose $N\geq 1$ such that $u_T\in \mathcal F^N T(V_{\omega \cdot 2})$.

Under the limit map projections $\ell_k \from V_{\omega \cdot 2} \to V_{\omega + k}$, we have that $U(\ell_N)(u) = \ell_{N+1}(m_{\omega \cdot 2}(u)) = w_{N+1}$.
We now project to the $T$-component of $V_{\omega + N + 1}$, giving $T(\ell_N)(u_T) = \operatorname{pr}_T(w_{N+1})$, and inspect where each side lies in the filtration.
On the one hand, since $T(\ell_N)$ preserves the filtration, the left-hand side lies in $\mathcal F^N T(V_{\omega + N})$.
On the other hand, by construction, the $T$-component of $w_{N+1}$ contains a nonzero term beginning with $v_{N+1}$.
This term is nonzero because $1_\Q \otimes z \neq 0$ and $\phi_N$ remains monic after tensoring with $\Q$.
Thus the right-hand side does not lie in $\mathcal F^N T(V_{\omega + N})$, giving the desired contradiction.

\end{proof}
\section{Convergence in $\omega\cdot(n+2)$ steps}
\label{sec:omega3}
\pratendSetLocal{category=omega3}
Fix $R$ to be a Noetherian ring of finite Krull dimension $n$. We shall give a convergence result for finitary functors on $\RMod$ via a filtration of $\RMod$ by Krull dimension.
The notion of Krull dimension extends naturally to finitely presented $R$-modules. Given such an $M$ we write $\dim M \coloneqq  \dim \left(R/\Ann M \right)$ and $\dim 0 =-\infty$, see \cite[Chapter 9]{eisenbud95}.  Note that we always have that $\dim M \leq n$.  The main reason dimension is useful is the following lemma.
\begin{lemma}\label{lem:extensionsmax}\cite[Lemma 10.62.7]{stacks-project}
Let $0 \to N \to M \to Q \to 0$ be a s.e.s. and $M$  be finitely generated. Then $\dim M  =  \max \left( \dim N,\dim Q \right)$.
\end{lemma}
This filtration on finitely presented modules induces a filtration on $\RMod$.
\begin{definition}
For $-1 \leq d \leq n$, let $\T_d \subseteq \RMod$ be the full subcategory generated under filtered colimits by  the finitely generated modules which have dimension at most $d$. For $M \in \RMod$ we write $t_d(M)$ for the directed colimit of all finitely presented submodules of $M$ that are in $\T_d$, i.e., the largest submodule of $M$ contained in $\T_d$.
\end{definition}
\begin{propositionE}[][all end]
The assignment $M \mapsto t_d(M)$ defines a functor.
\end{propositionE}
\begin{proofE}
Let $M \in \T_d$ be finitely generated and let $f \from M \to N$ be a morphism in $\RMod$.
Then we take its (epi, mono) factorisation  $f \from M \to \operatorname{Im}f \hookrightarrow N$. 
We have that $\operatorname{Im}f$ is a quotient of $M$ and$ \Ann M \subseteq \Ann \operatorname{Im}f$, which implies that $\dim \operatorname{Im}f \leq \dim M \leq d$. So $\operatorname{Im}f \subseteq t_d(N)$
The functoriality of $t_d$ then follows by restricting to each finitely generated submodule and taking the directed colimit to give the desired map $t_d(f) \from t_d(M) \to t_d(N).$
\end{proofE}
\begin{example}
\begin{enumerate}
\item
$\T_{-1}$ consists of the zero module and nothing else.
\item
$\T_0$ consists of the modules that are directed unions of submodules of finite length.
\item
If $R$ is a domain, then $\T_{n-1}$ is the category $ \RTors$ of torsion modules. Indeed a finitely generated module $M$ has $\dim M \leq n-1$ if and only if $\Ann M \neq 0$, which for a domain is precisely the condition that $M$ be torsion.
\item
In the case of $R = \Z$: $\T_0$ is the category $\ZTors$ and $\T_1 = \ZMod$. Given  a finitely presented $M = \Z^{\oplus m} \oplus \Z/p_1^{k_1}\Z \oplus \cdots \Z/p_r^{k_r}\Z $ we have that $t_0(M) \cong \Z/p_1^{k_1}\Z \oplus \cdots \Z/p_r^{k_r}\Z$  and $t_1$ is just the identity functor.
\item 
If $\dim R = n$, then $\T_n = \RMod.$
\end{enumerate}
\end{example}
\begin{textAtEnd}[category=omega3]
We can describe $t_d$ elementwise as follows via the following easy lemma.
\begin{lemma}
\label{lem:elementwise_td}
Let $x\in M$. Then $x \in t_d(M)$ if and only if $ \dim R/ \Ann{x} \leq d.$
\end{lemma} 
\begin{proof}
Suppose $x\in t_d(M).$ Then there is some finitely generated $N$ such that $x\in N$ and $ \dim R/ \Ann N \leq d.$   Then $\Ann N \subseteq \Ann (x)$ and so we have:
\begin{equation*}
\dim R / \Ann x \leq \dim R/ \Ann N \leq d.
\end{equation*}
where we have used the correspondence theorem for ideals in a quotient. This proves one of the directions. 

Conversely, if $ \dim R / \Ann{x} \leq d$, then $Rx$ is a finitely generated submodule of $M$ which has dimension less than or equal to $d$. The conclusion follows.
\end{proof}
\begin{lemma}
\label{lem:directed_colimits}
The functor $t_d$  preserves directed colimits.
\end{lemma} 
\begin{proof}
We need to show that the canonical map $ \colim_{i \in \mathcal I} t_d\left(M_i\right) \to t_d\left(\colim_{i \in \mathcal I} M_i \right) $ is an isomorphism in $\RMod$.
This is easy to see as by Lemma \ref{lem:elementwise_td}, $x \in t_d(M)$ if and only if $\dim R / \Ann x \leq d$.

Suppose $x \in t_d\left(\colim_{i \in \mathcal I} M_i \right)$.
Now, over a Noetherian ring all ideals are finitely generated, so $\Ann x = (r_1, \dots, r_n)$ for some $r_i \in R$.
We can choose a representative $x_k \in M_k$ of $x$ for some index $k$.
For each $j$, $r_j x_k$ is zero in the colimit and hence is zero at some later stage.
It follows that there exists an index $l \geq k$ such that all of $(r_1, \dots, r_n)$ annihilates the image $x_l$ of $x_k$ in $M_l$.
Thus $\Ann x \subseteq \Ann x_l$, so $\dim R / \Ann x_l \leq \dim R / \Ann x \leq d$ and hence $x_l \in t_d\left(M_l\right)$.
This proves surjectivity.

The canonical map is also monic because its composite with $t_d\left(\colim_{i \in \mathcal I} M_i \right) \hookrightarrow \colim_{i \in \mathcal I} M_i$ is monic, as directed colimits preserve monomorphisms.
\end{proof}
 \begin{proposition}
The category $\T_d$ is closed under submodules, quotients, extensions and direct sums. 
\end{proposition}
\begin{proof}
We first check the properties in the finitely generated case.
\emph{Submodules:} The annihilator of a submodule contains the annihilator of the module itself. 
By the correspondence theorem for ideals of quotient rings, it follows that the dimension of submodule must be less than or equal to the dimension of the module.

\emph{Quotients:} If an element $x\in \Ann M$ and $N$ is a submodule of $M$ then $x$ clearly annhilates $M/N.$ So the annihilator of a quotient module contains the annihilator of the original module.  Once again, by the correspondence theorem for ideals of quotient rings, it follows that the dimension of quotient must be less than or equal to the dimension of the module.

\emph{Extensions} This trivially follows from Lemma \ref{lem:extensionsmax}. Finite direct sums are a type of extension.

For arbitrary modules, we use that $M \in \T_d$ iff every fg. submodule of $M$ is of dimension at most $d$.
If $M \in \T_d$, each such submodule is finitely presented, so its inclusion factors through one stage of a filtered-colimit presentation of $M$.
The map into that stage is monic, so the finite case applies.
To see the converse, $M$ is the directed union of its finitely generated submodules.

\emph{Submodules:} Closure under submodules now follows immediately.

\emph{Quotients:} A finitely generated submodule of a quotient is the image of a finitely generated submodule of the original module, so closure under quotients follows from the finite case.

\emph{Extensions}  Finally, given a ses $0 \to M \to N \to L \to 0$ with $M, L \in \T_d$, every fg submodule $P \subseteq N$ fits into a ses
\begin{equation*}
0 \to P \cap M \to P \to \operatorname{Im}(P \to L) \to 0.
\end{equation*}
As $R$ Noetherian, all three modules are finitely generated.
The outer two have dimension at most $d$, so Lemma \ref{lem:extensionsmax} gives $\dim P \leq d$.
Thus $N \in \T_d$.
\end{proof}
\end{textAtEnd}
The subcategory $\T_d$ is closed under submodules, quotients and extensions.
It is also closed under arbitrary direct sums, since these are directed colimits of finite direct sums, which are extensions.
These properties make $\T_d$ a \emph{localizing subcategory} in the homological algebra literature \cite{gabriel62}.
\begin{propositionE}
The category $\T_d$ is a coreflective subcategory of $\RMod,$ with coreflector $t_d.$
\end{propositionE}
\begin{proofE}
Let $M\in \T_d$. The category $\T_d$ is closed under images as it is closed under quotients and, given a map $f \from M \to N$, we have that the map $M  \to \operatorname{Im}f$ is an epi, and therefore a quotient and therefore in $\T_d$. We thus obtain that $f$ factors uniquely though $i(N) \from t_d(N) \to N$ as required. 
\end{proofE}
We use throughout the remainder of this section that $t_d(N) = N \cap t_d(M)$ for submodules $N \subseteq M$. This follows from  closure under submodules and the maximality of $t_d$.
For each $-1 \leq d \leq n$, we define two classes of morphisms:
\begin{equation*}
  \mathcal E_d \coloneqq \setDef{f}{f \text{ is epi and } \ker f  \in \T_d},
  \qquad
  \M_d  \coloneqq \setDef{f}{t_d(\ker f) = 0}.
\end{equation*}
These form an orthogonal factorization system on $\RMod.$
In the $R = \Z$ case, $\M_0$ are the homomorphisms with torsion-free kernel, so they are monic on the torsion part.  Every monomorphism is in $\M_d$, and a map in both $\mathcal E_d$ and $\M_d$ must be an isomorphism. The extreme cases are easy to compute: $\M_{-1}$ consists of all morphisms and $\M_n$ is exactly the monomorphisms, as in $\RMod$ the monomorphisms are the maps with zero kernel. So therefore, one can think of the morphisms in between as being a weakened notion of a monomorphism:
\begin{equation*}
\M_{-1} \supseteq \M_0 \supseteq \M_1 \supseteq \cdots \supseteq \M_n
\end{equation*}
In other words, one should think of $\M_d$ as the maps that are \emph{monic up to $\T_d$.}
\begin{propositionE}[][all end]
\label{prop:ofs}
For each $d$, the pair $(\mathcal E_d, \M_d)$ is an orthogonal factorisation system on $\RMod$.
\end{propositionE}
\begin{proofE}
We may factor any morphism $f\from A \to B$ as
\begin{equation*}
  A \longrightarrow A/t_d(\ker f) \longrightarrow B,
\end{equation*}
where, by construction, the first arrow is in $\mathcal E_d$ and the second is in $\M_d$.
This latter fact follows from closure of $\T_d$ under extensions.
The kernel of the second arrow is $\ker f / t_d(\ker f)$.
If $Q \subseteq \ker f / t_d(\ker f)$ belongs to $\T_d$, then its inverse image in $\ker f$ is an extension of $Q$ by $t_d(\ker f)$ and so belongs to $\T_d$.
Using the maximality of $t_d(\ker f)$, it follows this inverse image must be equal $t_d(\ker f)$, and so $Q = 0$.
Thus $t_d(\ker f / t_d(\ker f)) = 0$, as required.

 To prove orthogonality, consider the following square
 \begin{equation*}
 \begin{tikzcd}
 A \rar{u} \dar{e} & X \dar{m}
 \\
 C \rar{v} & Y
 \end{tikzcd}
 \end{equation*}
 where $e\in \E_d$ and $m \in \M_d.$
Observe that  $m(u(\ker e)) = v(e(\ker e)) = v(0) = 0$ so we have $u(\ker e) \subseteq \ker m$.  Now $u(\ker e)\in \T_d$  so $u(\ker e) \subseteq t_d(\ker m) = 0$. Therefore, it follows from the usual (epi,mono)-factorisation system on $\RMod$ that $u$ factors as $u = w\circ e$ for a unique $w$. This gives us the desired lift as $m\circ w \circ e = m \circ u = v\circ e$ gives $m \circ  w = v$ as $e$ is an epi.

The classes obviously both contain all isomorphisms and are closed under composition.
\end{proofE}

To link this back to \cite{AdamekEA25}, observe that, by \cite[Rem.~2.6]{AdamekEA25}, a functor preserving binary intersections preserves monomorphisms. 
 Since $\M_n$ is the class of monomorphisms, Proposition \ref{prop:monopres} shows that the first condition of Theorem \ref{thm:omega3} implies monomorphism preservation.
Our  main convergence result is the following.
\begin{theorem}
\label{thm:omega3}
Let $R$ be a Noetherian ring with Krull dimension $n$ and let
$F\from \RMod \to \RMod$ be a finitary functor that
\begin{enumerate}
\item $F$ preserves countable directed intersections of monomorphisms;
\item $F$ preserves $\M_d$ for every $0\leq d \leq n-1$.
\end{enumerate}
Then $F$ has a terminal coalgebra obtained in $\omega\cdot (n+2)$ steps.
\end{theorem}
\begin{remark}
Condition (1) can be weakened to: $F$ preserves binary intersections and $F$ preserves the intersection of every descending chain of submodules of a finitely generated module, however this is less elegant to state.
\end{remark}
 Applied to $R = \Z$, where $n=1$, we obtain the following corollary.
\begin{corollary}
\label{thm:omega3Z}
Let $F\from \ZMod \to \ZMod$ be a finitary functor such that
\begin{enumerate}
\item $F$ preserves countable directed intersections of monomorphisms;
\item $F$ preserves maps that are monic when restricted to their torsion component.
\end{enumerate}
Then $F$ has a terminal coalgebra obtained in $\omega\cdot 3$ steps.
\end{corollary}
\begin{example}
Let $B$ be a projective $R$-module and $A$ arbitrary. The affine functor $F(X) = A \oplus (B \otimes X)$ is finitary, preserves intersections of countable families of subobjects, and preserves $\M_d$ for $-1 \leq d \leq n$. 
For the last claim, $\ker F(f) \cong B \otimes \ker f$ is a direct summand of a direct sum of copies of $\ker f$, since $B$ is projective.
Having zero $t_d$ is preserved by both operations.
It therefore satisfies the hypotheses of Theorem \ref{thm:omega3}. 
In particular, this applies to the linear weighted coalgebra functor $F(X) = \Z \oplus ( \Z^{L} \otimes X)$ as  $\Z^{L}$ is free for any finite set of labels $L$.
\end{example}
We conclude by explaining the proof of Theorem \ref{thm:omega3}. We summarise the intuition.
The categories $\RMod$ and $\T_n$ are a special case of a more general type of category called an \emph{abelian category}. 
Such categories possess a notion of quotient called the \emph{Serre quotient}. 
The category $\T_d / \T_{d-1}$  then satisfy the DCC-condition, and, in fact, the length of an equivalence class of a module $M$ within it can be precisely computed to be the following.
\begin{propositionE}
\label{lem:mu}
There is a function $\mu_d \from \T^{fp}_d \to \mathbb N$, where $\T^{fp}_d$ is the class of finitely presented modules contained in $\T_d$,  such that:
\begin{enumerate}
\item given a ses  $0\to M\to N \to L \to 0$ in $\T_d^{fp}$, we have $\mu_d(N) = \mu_d(M)+ \mu_d(L)$;
\item $\mu_d(N) = 0$ if and only if $N \in \T_{d-1}$.
\end{enumerate}
\end{propositionE}
\begin{proofE}
The proof is fairly standard commutative algebra, but requires some additional definitions to set up in a relatively self-contained fashion.
Let $\p$ be a prime ideal of $R$. 
The \emph{localization} $i\from R \to R_{\p}$ is characterized by the universal property that, for every ring morphism $f \from R \to S$, such that every $r \in R - \p$ is sent to an invertible element of $S$, there is a unique arrow such that following diagram commutes.
\begin{equation*}
\begin{tikzcd}
R \dar{i} \rar{f} & S
\\
R_\p \arrow[ur, dotted] & \
\end{tikzcd}
\end{equation*}
The example to keep in mind is $\Z_{(0)} \cong \Q,$ i.e., we add formal inverses to all elements of $\Z$ outside the ideal $(0)$.

Every morphism of rings $f\from R\to S$ induces an \emph{extension of scalars} functor $S \otimes_R - \from \RMod \to \Mod{(S)}$, which can also be thought of as tensoring by $S$ regarded as an $R$-module. 
In the special case of the localization functor $i\from R \to R_\p$, this  is denoted $(-)_\p \from \RMod \to \Mod{\left(R_\p\right)}$.
As we saw in Proposition \ref{silentlemma}, tensoring is a highly destructive operation: given a $\Z$-module $M$, one has that $M_{(p)}$ is $M$ with all torsion, except $p$-torsion, killed e.g., $(\Z/ 3\Z)_{(2)} = 0$ (and also with certain scalars inverted).
 We denote the set $\Supp M = \setDef{\p \subset R}{ \p \mbox{ prime and } M_\p \neq 0}$. 
It is well-known (and easily seen) that when $M$ is finitely generated these are the prime ideals such that $\Ann M \subseteq  \p.$

Let $N \in \T_d^{fp}$ for $0 \leq d \leq n$, and set
\begin{equation*}
  \mu_d(N) \coloneqq\ \sum_{\p  \mbox{ \small prime,} \dim R/\p  =  d}
     \operatorname{length}_{R_{\p}} \left( N_{\p} \right)
\end{equation*}
In the above the length of a module is the usual notion of longest strictly descending chain, see \cite[Section 10.52]{stacks-project} for the definition of length and its basic properties.

First, we check that the definition of $\mu_{d}$ is well-defined, that is, $\mu_d(N) < \infty$.
By definition, we have that $\dim N \leq d$, i.e., the longest descending chain of prime ideals in $R/\Ann N$ has length $d$ or less. 
Recall the basic fact that the prime ideals of $R/I$ are precisely the prime ideals of $R$ that contain $I.$ 
If $\p \in \Supp N$ and $\dim R / \p = d$, then there is a chain of length $d$, $\p_0\supsetneq\p_1\supsetneq\cdots\supsetneq\p_d=\p$, in $R$.
There cannot exist a prime $\p'$ with $\p\supsetneq\p'\supseteq\Ann N$ because then we would have a chain $\p_0 \supsetneq \p_1 \supsetneq \cdots \supsetneq \p_d=\p\supsetneq\p'$, and this would imply $\dim N \geq d+1$, which would contradict our starting assumption.
This implies that  $\p$ is \emph{minimal} in the set $\Supp N$ with respect to the partial order induced by set inclusion. 
There are finitely many such $\p$ by  \cite[Theorem 3.1]{eisenbud95}, and given such $\p$ one has $\operatorname{length}_{R_\p}(N_\p) < \infty$ for each such $\p$ by \cite[Corollary 2.18]{eisenbud95}. 
All summands indexed by primes outside $\Supp N$ vanish.
 So $\mu_d$ is well defined. 

Next we check the two conditions.
\begin{enumerate}
\item 
Firstly, the functor $(-)_\p$ preserves short exact sequences as it is equivalent to tensoring with $R_\p$ and this is a flat $R$ module: in other words $R_\p\otimes -$ preserves short exact sequences \cite[Proposition 2.5]{eisenbud95}. 
Secondly, length is additive \cite[Lemma 10.52.3]{stacks-project}, i.e., given a ses  $0\to M\to N \to L \to 0$, we have $\operatorname{length}(N) = \operatorname{length}(M)+ \operatorname{length}(L)$. It follows that $\mu_d$ is additive;
\item
Since the length of $N_\p$ is at least 1 for $\p \in \Supp N$, and $\Supp N$ is precisely the prime ideals containing $\Ann N$ (applying once again \cite[Corollary 2.7]{eisenbud95}), we have $\mu_d (N) = 0$ iff $\Supp N$ contains no prime
where $\dim R/\p=d$.
Since $\dim N \leq d$, this is equivalent to $\dim N \leq d-1$, and hence to $N \in \T_{d-1}$.
\end{enumerate}
\end{proofE}
\begin{example}
\label{ex:GKZMod}
Returning to $\ZMod$ case, the $\T_i$ filtration has two levels and there are hence two such functions to take care of $\mu_0 \from \ZTors ^{fp}\to \mathbb N$ and $\mu_1 \from \ZMod^{fp} \to \mathbb N$. 
The general form of a finitely presentable object is $M = \Z^{\oplus n} \oplus \Z/p_1^{k_1}\Z \oplus \cdots \Z/p_r^{k_r}\Z $ and
\begin{equation*}
\mu_0(t_0(M))= \sum_{i=1}^r k_i, \qquad \mu_1(M) = n.
\end{equation*}
Notice that $\mu_0$ can be interpreted as the length of the longest filtration of the torsion part of $M$ and $\mu_1$ is the longest filtration of the module $\mathbb Q \otimes M$ in $\mathbb Q\mathbf{Vect}.$ 
\end{example}
Next we consider the kernels $K_i$ of the limit maps $\ell_i\circ i_t$ out of a finitely presented submodule $i_t\from M_t\hookrightarrow L$, where $L$ is the limit of an inverse chain with connecting maps $a_i\in\M_{d-1}$.
Applying $t_d$ gives a descending chain $N_i\coloneqq t_d(K_i)$ of finitely presented modules
with intersection the zero module.
Since
\begin{equation*}
N_i/N_{i+1}\hookrightarrow K_i/K_{i+1}\hookrightarrow\ker a_i,
\end{equation*}
every nonzero quotient $N_i/N_{i+1}$ belongs to $\T_d$ but not to $\T_{d-1}$, and therefore has positive $\mu_d$.
We apply $\mu_d$ to this chain: by Proposition \ref{lem:mu}(1), every strict decrease strictly lowers the nonnegative integer $\mu_d(N_i)$, so the chain must eventually stabilise,
Its intersection is zero, so $t_d(K_s)=N_s=0$ for some $s$, and $\ell_j\circ i_t\in\M_d$ for every $j\geq s$ by the definition of $\M_d$.
We then use the finitary property of $F$ and its preservation of $\M_d$ to obtain the following lemma, which provides the inductive step of our main theorem.
\begin{lemmaE}
\label{lem:mdstep}
Let $F \from \RMod \to \RMod$ be finitary and suppose that it preserves $\M_d$ for $0\leq d\leq n$.
If, for some ordinal $\omega \cdot l$, the connecting maps $m_{\omega \cdot l + k }$ belong to $\M_{d-1}$ for every $k<\omega$, then $m_{\omega \cdot (l+1)}$ belongs to $\M_d$.
\end{lemmaE}
\begin{proofE}
This lemma follows by combining Lemmas \ref{prop:mdalpha} and  \ref{lem:mdchain}
\end{proofE}
We can now finally prove the main result.
\begin{proofsketch}
Proposition~\ref{prop:monopres} tells us our first assumption also ensures that $F$ also preserves the monomorphisms $\M_n$.

Starting with $\M_{-1}$, the class of all maps, by Lemma \ref{lem:mdstep}, in each $\omega$-block the connecting morphisms improve from being in $\M_{d-1}$ to being in $\M_d$ until we reach $\M_n$, the monomorphisms, after $n+1$ blocks of $\omega$ steps.
Here, preservation of $\M_d$ ensures that once
$m_{\omega\cdot(d+1)}\in\M_d$, all connecting maps in the
following block also belong to $\M_d$, since
$m_{\omega\cdot(d+1)+k} =F^k(m_{\omega\cdot(d+1)}).$
This allows us to repeat the argument in each block.

The final block is then a descending chain of submodules.
Preservation of its intersection (the first assumption of our theorem) means $FV_{\omega\cdot(n+2)}$ the limit of the shifted-by-one chain, and this must have the same limit as the original chain.
Thus the connecting map $m_{\omega\cdot(n+2)}$ is an isomorphism, giving the desired convergence at $\omega\cdot(n+2)$.
\end{proofsketch}
\begin{lemmaE}[][all end]
\label{prop:mdalpha}
Let $F \from \RMod \to \RMod$ be finitary and preserve $\M_d$, and let $\alpha$ be a limit ordinal such that every finitely presentable subobject $i_t \from M_t \hookrightarrow V_\alpha$ satisfies $\ell_k \circ i_t \in \M_d$ for some $k < \alpha$. 
Then the connecting morphism $m_\alpha \from FV_\alpha \to V_\alpha$ is in $\M_d$.
\end{lemmaE}
\begin{proofE}
Let $x\in t_d(\ker m_\alpha)$. We show that $x=0$.

First, express $V_\alpha$ as the directed colimit  $\colim_t M_t$ of its finitely presentable subobjects.
 As $F$ is finitary and $t_d$ preserves directed colimits, $t_d(FV_\alpha)=\colim_t t_d(FM_t)$, so $x=F(i_t)(y)$ for sufficiently large $t$ and some $y\in t_d(FM_t)$.
 
Choose any $j$ with $\ell_j \circ i_t \in \M_d$.
Then $F(\ell_j\circ i_t) \in \M_d$ because $F$ preserves $\M_d$. From the definition of $m_\alpha$, one has $F (\ell_j )= \ell_{j+1}\circ m_\alpha$, and we get that
\begin{equation*}
 F(\ell_j\circ i_t)(y) =  \ell_{j+1}\left(m_\alpha(x)\right) = 0 .
\end{equation*}
Since $y\in t_d(FM_t)$ and $F(\ell_j\circ i_t)(y)=0$, the identity $t_d(N)=N\cap t_d(M)$ for submodules gives $y\in t_d(\ker F(\ell_j\circ i_t))=0$. Hence $x=F(i_t)(y)=0$.
\end{proofE} 
\begin{lemmaE}[][all end]
\label{lem:mdchain}
Let $0\leq d \leq n$ and $(A_i, a_i)$ be an $\omega^{op}$-chain in $\RMod$ whose connecting morphisms $a_i \from A_{i+1} \to A_i$ are all in $\M_{d-1}$, with limit morphisms $\ell_i \from L \to A_i$. 
For every finitely presentable subobject $m \from M \rightarrowtail L$ there is a $k$ such that $\ell_k\circ m \in \M_d$.
\end{lemmaE}
\begin{proofE}
We prove this lemma by applying $\mu_d$ to the $\T_d$-parts of the kernels and showing that these are eventually zero.

Set $K_i = \ker(\ell_i \circ m) \subseteq M$. As $\ell_i = a_i \circ \ell_{i+1}$, we obtain a chain  of embeddings: $K_{i+1}\hookrightarrow K_i$. 
We can compute the limit of this chain, as the limit cone is jointly monic and $m$ is monic,  we have that $\bigcap_i K_i = 0$.
 Moreover, $\ell_{i+1}\circ m$ sends $K_i$ to $\ker a_i$ and the kernel of this map is precisely $K_{i+1}$. We thus obtain an inclusion:
\begin{equation*}
  K_i/K_{i+1} \hookrightarrow \ker a_i .
\end{equation*}
Now, we use the assumption that $a_i \in \M_{d-1}$. By definition, this implies that the module $\ker a_i$ has no nonzero submodule in $\T_{d-1}$, and thus $K_i/K_{i+1}$ does not either.

Next, we apply $t_d$ to the chain $K_i$, to obtain a chain  $N_i \coloneqq t_d(K_i).$ Note that we have the formula $N_i = N_0 \cap K_i$ so this is a chain of monos. 
By construction $\dim N_0 \leq d$, and the intersection can be computed as $\bigcap_i N_i \subseteq \bigcap_i K_i = 0$. 
Since $N_{i+1}=N_i \cap K_{i+1}$, the inclusion $N_i\subseteq K_i$ induces a monomorphism
\begin{equation*}
N_i/N_{i+1}\hookrightarrow K_i/K_{i+1}\hookrightarrow\ker a_i.
\end{equation*}
We have that $N_i/N_{i+1}$ is in $\T_d$ as it is a quotient of $N_i$. Furthermore, if $(N_i/N_{i+1} \neq 0$, then it is not in $\T_{d-1}$, since it embeds into $\ker a_i$, which has no nonzero submodule in $\T_{d-1}$.

Note that each $N_i$ is finitely presented, as it is a submodule of $M$ and $M$ is a finitely presented module over a Noetherian ring so all of its submodules are finitely presented.
So it makes sense to apply $\mu_d$ to them. 
 In particular it follows from the discussion above that if $N_i/N_{i+1} \neq 0$ then $\mu_d(N_i/N_{i+1})\geq 1$.

By Proposition \ref{lem:mu} we have $\mu_d(N_0) < \infty$, and by Proposition \ref{lem:mu}(i) applied to the short exact sequence
\begin{equation*}
0 \to N_{i+1}  \to  N_i \to  N_i/N_{i+1} \to 0
\end{equation*}
we get $\mu_d(N_i) - \mu_d(N_{i+1}) = \mu_d(N_i/N_{i+1})$.

From this, we can deduce that $N_{i+1} \subsetneq N_i$ forces $\mu_d(N_{i+1}) < \mu_d(N_i)$, and the chain $(\mu_d( N_i))_{i <\omega}$ cannot infinitely descend since it starts at  $\mu_d(N_0)$. So, it must be stationary from some $k>0$ on, whence $N_k = \bigcap_j N_j = 0$. Thus $t_d(K_k) = 0$ for all sufficently large $k$, i.e., $\ell_k \circ m \in \M_d$.
\end{proofE}

\section{Conclusion}
We have constructed a functor $U$ and shown that it does not converge in $\omega \cdot 2$ steps thus answering an open question of Ad\'{a}mek, Milius and Moss. 
We have given conditions for a finitary endofunctor on the category of $R$-modules, where the Krull dimension of a Noetherian $R$ is finite, to converge in countably many steps.

Several future directions remain.
Firstly, one might hope to find an analogue of Theorem \ref{thm:omega3} in semirings and semimodules.
A second direction for generalisation is to localising subcategories of abelian categories.
In order to combine these two, a generalisation of abelian categories, which can account for semimodules, has to be found.
Potentially, this may require developing the theory in terms of localisations of categories.
Finally, we think that our convergence results may provide insights into learning algorithms for automata.

\bibliographystyle{splncs04}
\bibliography{refs}

\begin{thebibliography}{10}
\providecommand{\url}[1]{\texttt{#1}}
\providecommand{\urlprefix}{URL }
\providecommand{\doi}[1]{https://doi.org/#1}

\bibitem{AdamekEA25}
Ad\'{a}mek, J., Milius, S., Moss, L.S.: {Terminal Coalgebras for Finitary
  Functors}. In: C\^{i}rstea, C., Knapp, A. (eds.) Proceedings of the 11th
  Conference on Algebra and Coalgebra in Computer Science. LIPIcs, vol.~342,
  pp. 3:1--3:16. Schloss Dagstuhl -- Leibniz-Zentrum f{\"u}r Informatik (2025).
  \doi{10.4230/LIPIcs.CALCO.2025.3}

\bibitem{AdamekEA19}
Ad\'{a}mek, J., Milius, S., Sousa, L., Wißmann, T.: Finitely presentable
  algebras for finitary monads. Theory Appl. Categ.  \textbf{34}(37),
  1179--1195 (2019)

\bibitem{ademek94}
Ad\'amek, J., Rosick\'y, J.: Locally presentable and accessible categories,
  London Mathematical Society Lecture Note Series, vol.~189. Cambridge
  University Press (1994). \doi{10.1017/CBO9780511600579}

\bibitem{Adamek03}
Adámek, J.: On final coalgebras of continuous functors. Theor. Comp. Sci.
  \textbf{294}(1),  3--29 (2003).
  \doi{https://doi.org/10.1016/S0304-3975(01)00240-7}

\bibitem{aristote26}
Aristote, Q., Petri\c~san, D.: Learning bottom-up tree automata valued in
  monoidal categories. In: Fo{SS}a{CS} 2026, Lecture Notes in Comput. Sci.,
  vol. 16503, pp. 21--42. Springer (2026). \doi{10.1007/978-3-032-22730-0\_2}

\bibitem{Basold-LinearCiruits}
Basold, H., Bonsangue, M., Hansen, H.H., Rutten, J.: ({{Co}}){{Algebraic
  Characterizations}} of {{Signal Flow Graphs}}. In: Horizons of the {{Mind}}.
  {{A Tribute}} to {{Prakash Panangaden}} - {{Essays Dedicated}} to {{Prakash
  Panangaden}} on the {{Occasion}} of {{His}} 60th {{Birthday}}. {{LNCS}},
  vol.~8464, pp. 124--145. Springer (2014). \doi{10.1007/978-3-319-06880-0_6}

\bibitem{Bhaduri23}
Bhaduri, P.: Coalgebras for bisimulation of weighted automata over semirings.
  Log. Methods Comput. Sci.  \textbf{19}(1),  Paper No. 4, 25 (2023).
  \doi{10.46298/lmcs-19(1:4)2023}

\bibitem{Bhattacharya15}
Bhattacharya, P.: The {$p$}-adic integers as final coalgebra. In: Logic,
  language, information, and computation. Lecture Notes in Comput. Sci.,
  vol.~9160, pp. 189--199. Springer (2015). \doi{10.1007/978-3-662-47709-0\_14}

\bibitem{bonchi12}
Bonchi, F., Bonsangue, M., Boreale, M., Rutten, J., Silva, A.: A coalgebraic
  perspective on linear weighted automata. Inform. and Comput.  \textbf{211},
  77--105 (2012). \doi{10.1016/j.ic.2011.12.002}

\bibitem{Boreale09}
Boreale, M.: Weighted bisimulation in linear algebraic form. In: C{ONCUR} 2009,
  Lecture Notes in Comput. Sci., vol.~5710, pp. 163--177. Springer, (2009).
  \doi{10.1007/978-3-642-04081-8_12}

\bibitem{dickson66}
Dickson, S.E.: A torsion theory for {A}belian categories. Trans. Amer. Math.
  Soc. pp. 223--235 (1966). \doi{10.2307/1994341}

\bibitem{dubut25}
Dubut, J.: Aczel-{M}endler bisimulations in a regular category. Log. Methods
  Comput. Sci.  \textbf{21}(3),  Paper No. 23, 56 (2025).
  \doi{10.46298/lmcs-21(3:23)2025}

\bibitem{eilenberg74}
Eilenberg, S.: Automata, Languages, and Machines A, Pure and Applied
  Mathematics, vol.~59. Elsevier (1974).
  \doi{https://doi.org/10.1016/S0079-8169(08)60890-9}

\bibitem{eisenbud95}
Eisenbud, D.: Commutative algebra, Graduate Texts in Mathematics, vol.~150.
  Springer-Verlag (1995). \doi{10.1007/978-1-4612-5350-1}

\bibitem{gabriel62}
Gabriel, P.: Des cat\'egories ab\'eliennes. Bull. Soc. Math. France
  \textbf{90},  323--448 (1962),
  \url{http://www.numdam.org/item?id=BSMF_1962__90__323_0}

\bibitem{HKR14:SDE}
Hansen, H.H., Kupke, C., Rutten, J.: Stream {{Differential Equations}}:
  {{Specification Formats}} and {{Solution Methods}}. Logical Methods in
  Computer Science  \textbf{13}(1) (2017). \doi{10.23638/LMCS-13(1:3)2017},
  \url{https://doi.org/10.23638/LMCS-13(1:3)2017}

\bibitem{rot20}
van Heerdt, G., Kupke, C., Rot, J., Silva, A.: Learning weighted automata over
  principal ideal domains. In: Fo{SS}a{CS} 2020, Lecture Notes in Comput. Sci.,
  vol. 12077, pp. 602--621. Springer (2020).
  \doi{10.1007/978-3-030-45231-5\_31}

\bibitem{isaksen02}
Isaksen, D.C.: A cohomological viewpoint on elementary school arithmetic. Amer.
  Math. Monthly (9),  796--805 (2002). \doi{10.2307/3072368}

\bibitem{leinster11}
Leinster, T.: A general theory of self-similarity. Adv. Math.  \textbf{226}(4),
   2935--3017 (2011). \doi{10.1016/j.aim.2010.10.009}

\bibitem{MacLane98}
MacLane, S.: Categories for the working mathematician. Springer (1998).
  \doi{10.1007/978-1-4757-4721-8}

\bibitem{Milius10:SoundCompleteCalculus}
Milius, S.: A {{Sound}} and {{Complete Calculus}} for {{Finite Stream
  Circuits}}. In: Proceedings of {{LICS}} 2010. pp. 421--430 (2010).
  \doi{10.1109/LICS.2010.11}, \url{https://doi.org/10.1109/LICS.2010.11}

\bibitem{Rutten00}
Rutten, J.: Universal coalgebra: a theory of systems. Theor. Comp. Sci.
  \textbf{249}(1),  3--80 (2000).
  \doi{https://doi.org/10.1016/S0304-3975(00)00056-6}

\bibitem{Rutten05:CoindCalcStreams}
Rutten, J.: A coinductive calculus of streams. Mathematical Structures in
  Computer Science  \textbf{15}(1),  93--147 (2005).
  \doi{10.1017/S0960129504004517},
  \url{http://dx.doi.org/10.1017/S0960129504004517}

\bibitem{ruttenTutorialCoinductiveStream2005}
Rutten, J.J.M.M.: A tutorial on coinductive stream calculus and signal flow
  graphs. Theor. Comput. Sci.  \textbf{343}(3),  443--481 (2005).
  \doi{10.1016/j.tcs.2005.06.019},
  \url{https://doi.org/10.1016/j.tcs.2005.06.019}

\bibitem{schmid25}
Schmid, T., Noquez, V., Moss, L.S.: Fractals from regular behaviours. Log.
  Methods Comput. Sci.  \textbf{21}(2),  Paper No. 25, 45 (2025).
  \doi{10.46298/lmcs-21(2:25)2025},
  \url{https://doi.org/10.46298/lmcs-21(2:25)2025}

\bibitem{stacks-project}
{Stacks project authors}, T.: The {Stacks project}.
  \url{https://stacks.math.columbia.edu} (2026)

\bibitem{Staton2011-Bisim}
Staton, S.: Relating coalgebraic notions of bisimulation. Logical Methods in
  Computer Science  \textbf{7}(1),  1--21 (2011).
  \doi{10.2168/LMCS-7(1:13)2011}

\bibitem{urbat20}
Urbat, H., Schr\"oder, L.: Automata learning: an algebraic approach. In:
  Proceedings of the 35th {A}nnual {ACM}/{IEEE} {S}ymposium on {L}ogic in
  {C}omputer {S}cience ({LICS} 2020). p.~15. ACM (2020).
  \doi{10.1145/3373718.3394775}

\bibitem{weibel94}
Weibel, C.A.: An introduction to homological algebra, Cambridge Studies in
  Advanced Mathematics, vol.~38. Cambridge University Press (1994).
  \doi{10.1017/CBO9781139644136}

\bibitem{Worrell99}
Worrell, J.: Terminal sequences for accessible endofunctors. Electr. Notes in
  Theor. Comp. Sci.  \textbf{19},  24--38 (1999).
  \doi{https://doi.org/10.1016/S1571-0661(05)80267-1},
  \url{https://www.sciencedirect.com/science/article/pii/S1571066105802671}

\bibitem{Worrell05}
Worrell, J.: On the final sequence of a finitary set functor. Theor. Comp. Sci.
   \textbf{338}(1),  184--199 (2005). \doi{10.1016/j.tcs.2004.12.009}

\end{thebibliography}
\newpage
\appendix
\section{Details on~\cref{sec:prelim}}
\printProofs[prelim]
\section{Details on~\cref{sec:counterexample}}
\printProofs[counterexample]
\section{Details on~\cref{sec:omega3}}
\printProofs[omega3]
\begin{proof}[Proof of Theorem \ref{thm:omega3}]
By Proposition \ref{prop:monopres}, the first condition implies that $F$ preserves $\M_n$.
Together with the second condition, this gives preservation of $\M_d$ for every $0 \leq d \leq n$.
We shall show by induction, that for every $0\leq d\leq n$ the morphism
\begin{equation*}
m_{\omega\cdot (d+1)}\from FV_{\omega\cdot (d+1)} \longrightarrow V_{\omega\cdot(d+1)}
\end{equation*}
is in $\M_d$.

For the base case $d = 0$,  we apply Lemma \ref{lem:mdchain} to the chain $(V_k, m_k)_{k<\omega}$. 
The class $\M_{-1}$ is all morphisms so, in particular, the morphisms in the chain. This satisfies the hypothesis of Proposition \ref{prop:mdalpha} at $\alpha = \omega$ and $d = 0$.
 As $F$ preserves $\M_0$, we may conclude that $m_\omega \in \M_0$.

The inductive step is exactly the same: let $1 \leq d \leq n$ and suppose $m_{\omega\cdot d} \in \M_{d-1}$. 
As all the other morphisms in the block are of the form $m_{\omega\cdot d + k} = F^k(m_{\omega\cdot d})$, they are also in $\M_{d-1}$, since $d-1\leq n-1$ and $F$ preserves $\M_{d-1}$ by hypothesis.
 Hence the block
\begin{equation*}
V_{\omega\cdot d} \longleftarrow V_{\omega\cdot d + 1}\longleftarrow V_{\omega\cdot d + 2}\longleftarrow \cdots
\end{equation*}
is an $\omega^{op}$-chain with all connecting morphisms in $\M_{d-1}$ and limit $V_{\omega\cdot(d+1)}$, so Lemma \ref{lem:mdchain} applies and verifies the hypothesis of Proposition \ref{prop:mdalpha} at $\alpha = \omega\cdot (d+1)$.
 As $F$ preserves $\M_d$, when $d= n$ this is implied by Proposition \ref{prop:monopres}, we conclude $m_{\omega\cdot(d+1)} \in \M_d$, which completes the induction.
Taking $d = n$ and recalling that $\M_n$ is the class of monomorphisms, we obtain that $m_{\omega\cdot(n+1)}$ is monic, and hence, since  $F$ preserves monos by Proposition \ref{prop:monopres}, that every connecting morphism of the block
\begin{equation*}
V_{\omega\cdot(n+1)} \longleftarrow V_{\omega\cdot(n+1) + 1}\longleftarrow V_{\omega\cdot(n+1) + 2}\longleftarrow \cdots
\end{equation*}
is monic. 
We now apply \cite[Proposition 3.1]{AdamekEA25} to this $\omega^{op}$-chain: its hypotheses hold by the previous sentence, by the fact $F$ preserves monos, by Proposition \ref{prop:monopres} and by (1) respectively. 
We conclude that $F$ preserves the limit $V_{\omega\cdot(n+1)+\omega} = V_{\omega\cdot(n+2)}$.
Therefore $\nu F = V_{\omega\cdot(n+2)}$. This concludes the proof.
\end{proof}
\end{document}